\documentclass[journal]{IEEEtran}

\usepackage[utf8]{inputenc}
\usepackage{cite}
\usepackage{setspace}
\usepackage{epsf}\epsfverbosetrue
\usepackage{graphics,epsfig}
\usepackage{graphicx,epsfig}
\usepackage{epsfig}
\usepackage{multirow}
\usepackage{alltt}
\usepackage{subfigure}

\usepackage{mathtools}

\usepackage{tcolorbox}
\usepackage{float}
\usepackage{url}
\usepackage{graphicx}
\usepackage{verbatim} 
\usepackage{footnote} 
\usepackage{latexsym} 
\usepackage{sidecap} 
\usepackage{wrapfig}
\usepackage{eso-pic}
\usepackage{fix-cm}
\usepackage{algorithm}
\usepackage{algpseudocode}
\usepackage{color}
\usepackage{wrapfig}
\usepackage{multicol}
\usepackage{flowchart}
\usetikzlibrary{shapes,arrows,matrix,decorations.pathreplacing,shapes.geometric,positioning,calc}

\usepackage{layout}

\usepackage{amsmath}
\usepackage[
  locale = DE 
]{siunitx}

\usepackage{textcomp}
\usepackage{multirow}
\usepackage{mathtools}
\usepackage{soul} 
\usepackage{amsmath} 
\usepackage{verbatim}
\usepackage{csvsimple} 
\usepackage{array}
\usepackage{subfig}
\usepackage[normalem]{ulem}
\usepackage{booktabs}
\newcolumntype{P}[1]{>{\RaggedRight\arraybackslash}p{#1}}
\newcommand{\tabitem}{\textbullet~~}

\usepackage{pifont}
\newcommand{\tick}{\ding{52}}%
\newcommand{\cross}{\ding{55}}%
\newcommand{\mult}{\ding{93}}%

\newcommand{\comst}[1]{\textcolor{black}{#1}}
\newcommand{\mubcom}[1]{\textcolor{black}{#1}}

\newcommand{\rev}[1]{\textcolor{black}{#1}}

\newcommand{\revtwo}[1]{\textcolor{black}{#1}}

\newcommand{\final}[1]{\textcolor{black}{#1}}

\begin{document}

\title{Anomaly Detection in Blockchain Networks: A Comprehensive Survey}

\author{Muneeb Ul Hassan, Mubashir Husain Rehmani, and Jinjun Chen
\thanks{M. Ul Hassan and J. Chen are with the Swinburne University of Technology, Hawthorn VIC 3122, Australia  (e-mail:  muneebmh1@gmail.com; jinjun.chen@gmail.com).}
\thanks{M.H. Rehmani is with the Department of Computer Science, Munster Technological University, Ireland (e-mail: mshrehmani@gmail.com).}
}

\maketitle

\begin{abstract}
Over the past decade, blockchain technology has attracted a huge attention from both industry and academia because it can be integrated with a large number of everyday applications of modern information and communication technologies (ICT). Peer-to-peer (P2P) architecture of blockchain enhances these applications by providing strong security and trust-oriented guarantees, such as immutability, verifiability, and decentralization. \rev{Despite these incredible features} that blockchain technology brings to these ICT applications, \final{recent research} has indicated \final{that the strong} guarantees are not sufficient enough and blockchain \rev{networks may still be prone to various} security, privacy, and reliability issues. In order to overcome these issues, it is important to identify the anomalous behaviour within the actionable time frame. In this article, we provide an in-depth survey regarding integration of anomaly detection models in blockchain technology. \rev{For this, we first discuss how anomaly detection can aid in ensuring security} of blockchain based applications. Then, we demonstrate certain fundamental \final{evaluation metrics} and key requirements that can play a critical role while developing anomaly detection models for blockchain. \rev{Afterwards, we present a thorough survey of various anomaly detection models from the perspective of each layer of blockchain}. Finally, we conclude the article by highlighting certain \rev{important challenges alongside discussing how they can serve as future research directions for new researchers in the field.}

\end{abstract}



\tikzstyle{decision} = [diamond, draw, fill=blue!50]
\tikzstyle{line} = [draw, -stealth, line width= 0.4mm]
\tikzstyle{elli}=[draw, ellipse, fill=red!50,minimum height=5mm, text width=5em, text centered]
\tikzstyle{firstblock} = [draw, rectangle, fill=orange!60, rounded corners, minimum height= 10mm, text width=5.5em, minimum width = 20mm, text width=11em, text centered]

\tikzstyle{secondblock} = [draw, rectangle, fill=blue!20, rounded corners, minimum height= 7mm, text width=3.5em, text centered]

\tikzstyle{thirdblock} = [draw, rectangle, fill=orange!70, rounded corners, minimum height= 7mm, text width=3.5em, text centered]

\tikzstyle{fourthblock} = [draw, rectangle, fill=yellow!65, rounded corners, minimum height= 7mm, text width=3.5em, text centered]

\tikzstyle{fifthblock} = [draw, rectangle, fill=lime!60, rounded corners, minimum height= 7mm, text width=5.5em, text centered]


\section{Introduction}

\rev{The tremendous success of Bitcoin as a cryptocurrency grabbed the attention of \final{researchers who} then started to explore the underlying technology behind this cryptocurrency named blockchain.} Research works investigated \final{that blockchain} integrated with information and communication technologies (ICT) have a vast number of use cases in daily life applications, such as supply chain, finances, IoT operations, cloud services, etc. Since then, applications of blockchain technology are being explored by researchers and this number is continuously increasing~\cite{introanomaly02}.\\
 From a technological point of view, blockchain is an immutable append-only chain of blocks which works over decentralized peer-to-peer (P2P) network~\cite{introanomaly03}. \rev{Each block in the decentralized ledger \rev{is composed of two critical components, \final{one is transactions and} the other is block header.} Transactions can be termed as a validated set of records between \final{peers, which can} be of multiple types ranging from financial transactions to real-time \final{reported} data. \rev{The second critical component of the block is its header, which \final{is comprised of} critical information,} such as nonce, block hash, previous block hash, time stamp, Merkle root, etc. Since each block is linked with each other with a secure hash function, thus, this property of the blockchain makes the complete network secure and immutable~\cite{introanomaly08}. \rev{Similarly, from a functioning perspective, blockchain can} further be classified into multiple layers named as data layer, network layer, incentive layer, and smart contract layer. Each layer in the blockchain taxonomy has its own functionalities and responsibilities. A detailed discussion about functioning and layer oriented architecture of blockchain has been \final{provided later} in this article (Section~\ref{BlockLayers}.).}\\
\rev{Despite these benefits, blockchain technology is not 100\% secure} and it is still prone to certain attacks and issues~\cite{contract06, introanomaly09}. \rev{Usually, the purpose of launching the attacks over the blockchain network is the capital or popularity of the system. For example, a successful attack on the Bitcoin/Ethereum network can result in reduction of its market value price. A very similar attack was carried out in 2017 over the Bitcoin network, when the network was flooded with a huge number of spam transactions \final{in order to cause delay and stall the transaction verification process, which in turn will increase} the fee of mining process of Bitcoin. This delay in the transaction \final{process resulted in delayed payments of approximately} \$700million USD~\cite{comstref01}. Similarly, attackers also carry out attacks on a decentralized application or cryptocurrency to reduce its popularity. For instance,  in 2017, the exchange Bitfinex faced a distributed denial-of-service (DDoS) attack in the network, which caused the suspension of service temporarily~\cite{survey09}.  Similar to these attacks, a large number of Ponzi schemes over the blockchain network have been developed in order to steal money from legitimate users, which attract layman users by promising future incentives, etc. \final{Alongside these,} a large number of malicious accounts are being created regularly on cryptocurrencies in order to carry out money laundering. Moreover, in certain anomalous attacks, malicious forks are created and deployed to overcome computational power and to carry out double spending in the network~\cite{introanomaly07}. Thus, in order to get most out of the blockchain network, it is equally important to detect the occurrence of these attacks and vulnerabilities in a timely manner.}\\


\begin{figure}
\centering
  \includegraphics[scale=0.36]{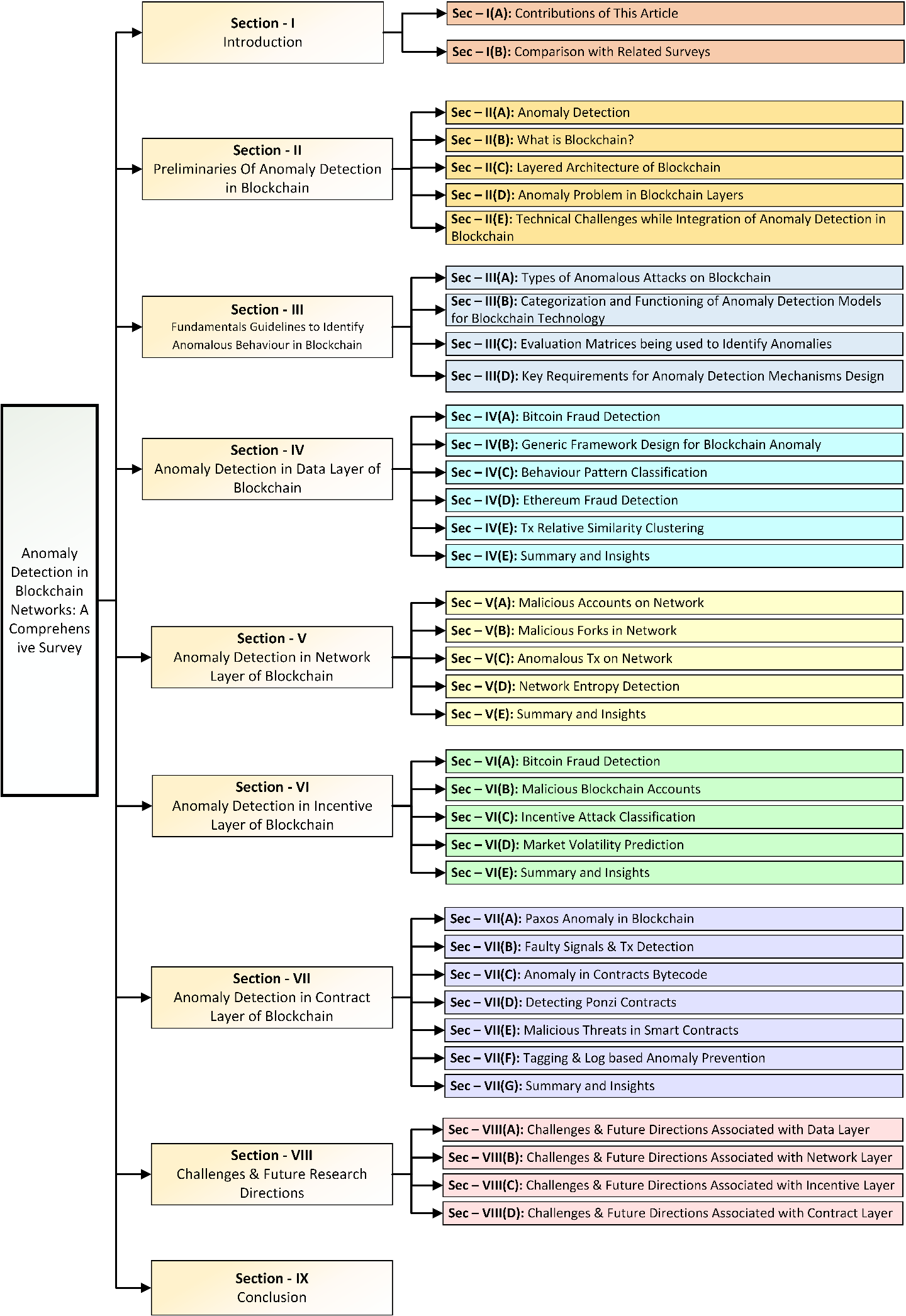}
  \caption{\rev{An Illustrative Overview of Structure of Our Survey Article. }}
  \label{fig:structure}
\end{figure}


To provide the successful detection and prediction of such attacks over blockchain, \rev{the field of anomaly detection for blockchain comes into action.} The major functionality is to detect or even predict the future occurrence of any vulnerability in the network in order to take a timely action against them. A large number of anomaly detection models are being created and deployed by researchers for \final{various blockchain networks.} From a generic point of view, these models can be categorized on the basis of different layers of blockchain. E.g., certain models work over prediction of anomalous commands in the smart contracts, \rev{so these models come under anomaly detection for smart contracts.} Similarly, some models work over detection of malicious block deployment, so these models can be categorized under the umbrella of \rev{anomaly detection in the data layer. Overall, it is important to mention anomaly detection is one of the important fields in order to secure the future blockchain network, and a vast amount of work is being carried out in this field from various perspectives which we will discuss in this survey.} \rev{Moreover, an illustrative organizational structure for our survey article has been presented in Fig.~\ref{fig:structure}.}

\begin{table*}[htbp]
 \centering
 \scriptsize
 \captionsetup{labelsep=space}
 \captionsetup{justification=centering}
 \caption{\textsc{\\Comparison of Existing Survey Works in the Similar field of Blockchain and Anomaly Detection with their Major Contribution and Certain Scopes. \newline} \footnotesize \textbf{Acronyms:} Classification of Blockchain Anomalies (CoBA), Anomaly in Contract Layer (AiCL), Anomaly in Data Layer (AiDL), Anomaly in Network Layer (AiNL), Anomaly in Incentive Layer (AiIL), Existing Works in Blockchain Anomaly Detection (EWiBAD).  Tick(\tick)~Shows that the mentioned topic is covered, Cross(\cross)~shows that the provided domain is not covered, and Asterisk(\mult)~shows that the particular topic is partially covered.}
    \begin{tabular}{|p{1.7em}|p{2em}|p{27em}|p{2em}|p{2em}|p{2em}|p{2em}|p{2em}|p{4em}|}
    \hline
    &  & &  \multicolumn{5}{c}{\centering  \bfseries ~~~~~~~~Scope} &\\
\cline{4-9}
\rule{0pt}{2ex}
    \centering \textbf{Ref No.}  & \centering \textbf{Year} & \centering \textbf{Major Contribution of Surveys} & \textbf{CoBA} &  \textbf{AICL} & \textbf{AIDL} & \textbf{AiNL} & \textbf{AiIL} & \textbf{EWiBAD} \\
    \hline
    
    \rule{0pt}{2ex}
    \cite{survey01} & 2018  & A comprehensive survey about security concerns and their countermeasures in Bitcoin. & \cross  & \cross & \cross & \mult & \mult & \mult \\
    \hline
    
    \rule{0pt}{2ex}
    ~\cite{survey02} & 2018  & A thorough literature about privacy and anonymity of Bitcoin-like systems. & \cross  & \cross & \cross & \cross & \mult & \mult \\
    \hline
    
    \rule{0pt}{2ex}
    ~\cite{survey14} & 2018  & A detailed study about integration of intrusion detection systems with blockchain technology. & \cross  & \cross & \cross & \mult & \cross & \cross \\
    \hline
    
    \rule{0pt}{2ex}
    ~\cite{survey03} & 2019  & A detailed survey of blockchain-based works in various security services. & \cross  & \cross & \cross & \cross & \cross & \cross \\
    \hline
    
    \rule{0pt}{2ex}
    ~\cite{survey04} & 2019  & A comprehensive discussion about privacy issued in IoT scenarios operating on blockchain. & \cross  & \cross & \cross & \cross & \cross & \cross \\
    \hline
    
    \rule{0pt}{2ex}
    ~\cite{survey05} & 2019  & A thesis works evaluating various anomaly detection mechanisms in blockchain. & \mult  & \cross & \mult & \cross & \mult & \mult \\
    \hline
    
    \rule{0pt}{2ex}
    ~\cite{survey06} & 2019  & A survey on integration strategies of blockchain technology with IoT and beyond from application perspective. & \cross  & \cross & \cross & \cross & \cross & \cross \\
    \hline
    
    \rule{0pt}{2ex}
    ~\cite{survey07} & 2019  & A comprehensive survey about blockchain, its functioning, and applicability in various scenarios. & \cross  & \cross & \cross & \cross & \cross & \cross \\
    \hline
    
    \rule{0pt}{2ex}
    ~\cite{survey16} & 2019  & A survey over security assurance and correction verifications of smart contracts deployed on blockchain. & \cross  & \cross & \cross & \cross & \cross & \cross \\
    \hline
    
    \rule{0pt}{2ex}
    ~\cite{survey08} & 2020  & A thorough survey on security attacks and their countermeasures for blockchain based IoT and IIoT systems. & \cross  & \cross & \cross & \cross & \cross & \cross \\
    \hline
    
    \rule{0pt}{2ex}
    ~\cite{survey09} & 2020  & Worked over exploration of attack surfaces and attacks vectors in blockchain environment. & \cross  & \cross & \cross & \cross & \cross & \cross \\
    \hline
    
    \rule{0pt}{2ex}
    ~\cite{survey10} & 2020  & A detailed investigation of blockchain from perspective of security reference architecture for blockchain. & \cross  & \cross & \cross & \cross & \cross & \cross \\
    \hline
    
    \rule{0pt}{2ex}
    ~\cite{survey11} & 2020  & A book chapter over anomaly detection approaches in blockchain technology. & \tick & \mult & \mult & \cross & \mult & \mult \\
    \hline
    
    \rule{0pt}{2ex}
    ~\cite{survey12} & 2020  & A detailed investigation of integration of privacy preservation via differential privacy strategy in blockchain technology. & \cross  & \cross & \cross & \cross & \cross & \cross \\
    \hline
    
    \rule{0pt}{2ex}
    ~\cite{survey13} & 2020  & A survey over necessities of privacy services, security issues, and applications of blockchain technology. & \cross  & \cross & \cross & \cross & \cross & \cross \\
    \hline
    
    \rule{0pt}{2ex}
    ~\cite{introanomaly06} & 2020  & A comprehensive survey of consensus algorithms in blockchain based systems. & \cross  & \cross & \cross & \cross & \cross & \cross \\
    \hline
    
    \rule{0pt}{2ex}
    ~\cite{survey17} & 2020  & A thorough technical overview of blockchain smart contracts.  & \cross  & \cross & \cross & \cross & \cross & \cross \\
    \hline
    
    \rule{0pt}{2ex}
    ~This Work & 2021  & A comprehensive survey on anomaly detection in blockchain technology from perspective of identification, integration, requirement, and methodologies for anomaly detection in blockchain. & \tick & \tick & \tick & \tick & \tick & \tick \\
    \hline
    \end{tabular}%
  \label{tab:comparison}%
\end{table*}%

\subsection{Contributions of This Article}
\rev{Certain surveys have been published in the field of blockchain, however,} to the best of our knowledge, none of them provides an in-depth overview of anomaly detection in blockchain technology. To summarize, the major contributions of our work are as follows:
\begin{itemize}
\item \rev{We work over providing generalist audience a brief overview regarding the field of anomaly detection in blockchain}
\item \mubcom{We provide detailed discussion about classification of anomalous attacks, their detection models, and \final{the} existing works in blockchain technology alongside providing some fundamental \final{metrics and} key requirements for their robust and timely identification.}
\item We highlight critical challenges in blockchain based anomaly detection that needs to be solved alongside providing a brief overview of future directions associated with these challenges.
\end{itemize}

\subsection{Comparison with Related Surveys}

\rev{The concept of anomaly detection for conventional networks has been in discussion in industry and academia \final{for a long time,} and a plethora of work has been carried over it so far. However, traditional anomaly detection models cannot directly be integrated with blockchain technology because blockchain has multiple aspects which makes it distinct from traditional networks, such as consensus mechanism, smart contract, decentralized P2P transactions, lack of central authority, etc. \final{Therefore,} in the presence of all these features, traditional anomaly detection models cannot be applied directly to blockchain. Thus, there is a dire need to develop anomaly detection models purely for blockchain based networks and applications.} \\
Our survey on anomaly detection \rev{in blockchain networks is distinctive from all past surveys because we cover the aspect of anomaly detection in detail from basics to integration perspective at different layers of blockchain technology. A table comprising of comparison of our proposed work with other similar works has been given in Table.~\ref{tab:comparison}.} Comparatively, in literature, a detailed survey discussing various security threats and their machine-learning based countermeasures \final{for Bitcoin has} been presented by Mohamed~\textit{et al.} in~\cite{survey01}.Similarly, another similar survey discussing the privacy and anonymity of Bitcoin and similar cryptocurrencies has been presented by authors in~\cite{survey02}. \rev{A short review discussing intrusion detection systems (IDS) in blockchain technology has been presented by Meng~\textit{et al.} in~\cite{survey14}. Similar to this, another survey work \final{which emphasizes the use of blockchain as a critical tool for security services has been presented by Salman~\textit{et al.} in~\cite{survey03}.}}\\ 
\rev{A work discussing privacy preservation strategies in blockchain based Internet of Things (IoT) systems has been presented by authors in~\cite{survey04}.} The next work~\cite{survey05} is a master’s thesis rather than a survey work, which discussed the implementation of \rev{seven anomaly detection models in blockchain scenarios}. Moving towards generic surveys of blockchain, a very comprehensive survey \final{which discusses} blockchain from theory to application perspective of IoT have been presented by Wu~\textit{et al.} in~\cite{survey06}. Another similar survey providing a thorough discussion about blockchain, and its useful scenarios have been presented by authors in~\cite{survey07}. \\
From security perspective, a short article discussing security verification have been presented by Liu~\textit{et al.} in~\cite{survey16}. Two other articles on very similar topic of security and attacks of blockchain \final{have} also been presented by authors in~\cite{survey08, survey09}. Similarly, a comprehensive work providing details about threats, vulnerabilities, and resilience models from security perspective of blockchain have been presented by authors in~\cite{survey10}. Alongside this, a very short work discussing anomaly detection in blockchain via data mining methods have been presented by authors in~\cite{survey11}. Apart from anomaly, a work providing an in-depth evaluation and discussion about integration of differential privacy mechanism in layers of blockchain \final{technology has} been presented by authors in~\cite{survey12}. Another work covering a brief overview of security and privacy threats alongside future applications have been published by authors in~\cite{survey13}. Distinct from these threats, a detailed survey providing an overview of consensus models in blockchain networks have been given in~\cite{introanomaly06}. Continuing this trend, a detailed survey covering advances in the field of blockchain based smart contracts have been published by Kemmoe~\textit{et al.} in~\cite{survey17}.\\

\section{Preliminaries of Anomaly Detection in Blockchain}

\begin{figure*}
\centering
  \includegraphics[scale=0.15]{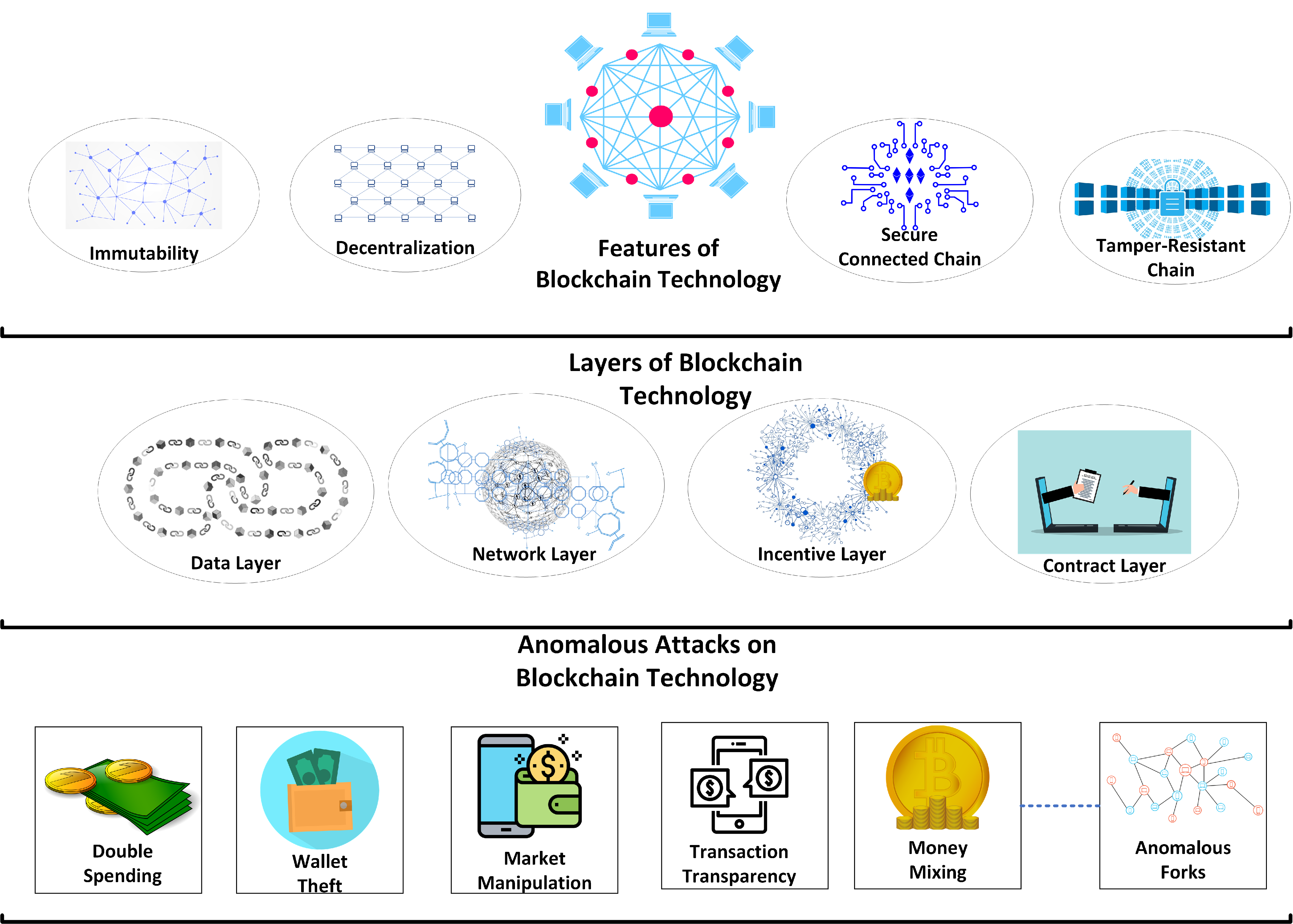}
  \caption{\rev{An Illustration of Blockchain Features, Blockchain Layers, and Certain Anomalous Attacks on Blockchain. }}
  \label{fig:blockchain}
\end{figure*}

\subsection{What is Blockchain?}

\subsubsection{Development Phases of Blockchain}
\rev{Blockchain technology has gone through a process of transformations with time. For instance, the first era of blockchain also named Blockchain 1.0 started in \final{2008 when} Satoshi Nakamoto introduced the notion of a decentralized consensus via proof of work, which is resistant to double spending and can successfully validate transactions without the help of any centralized authority. In this era, a large number of cryptocurrencies  such as Bitcoin, Ethereum, Ripple, etc. got formed. The first phase of blockchain enhanced the notion of decentralization, but it was just limited to financial transactions. Therefore, to improve it further, in the second era of blockchain (known as Blockchain 2.0), researchers worked over integration of feature of programmability in the network, with the help of smart contracts. This started with the development of basic DApps by Ethereum 2.0, which provided users with an environment to code. In this way, blockchain users were able to develop and write intelligent smart contracts according to \final{their requirements.} The third era of blockchain is currently ongoing and rapid development in the applicability of blockchain can be seen around us. For example, blockchain is being used with a diverse number of everyday applications, such as property management, banking, education, healthcare, energy systems, vehicular networks, etc.} 

\subsubsection{Functioning of Blockchain}
\rev{As the name suggests, blockchain is a chain of blocks, which are connected together by strong cryptographic guarantees. \rev{Each block has two parts, one is the header and the other part is the body of the block.} Header contains all linking information, such as hash of its body contents, hash of previous block, time-stamp, etc. \rev{Moreover, the body of the block comprises all the transactions. Since the hash content of a block is calculated and stored inside the block, therefore, tampering with anything in the block body will change the hash of the header,} \final{and as a result} this block will not be linked with the next block in the chain.} In this way, blockchain technology ensures that the ledger remains tamper-proof. Alongside this, each of the blockchain nodes has a copy of this digital ledger, thus, a single node cannot play with the complete chain as well, which ensures that the ledger is immutable~\cite{survey12}. All the nodes of blockchain operate in decentralized manner and have strong cryptographically secured communication between them which obeys the rules of P2P networking. All these functionalities combine to form a modern day technology, which is named as blockchain. \rev{A graphical illustration of features of blockchain alongside an illustration of layers and anomalies of blockchain has been presented in Fig.~\ref{fig:blockchain}.}

\subsection{Layered Architecture of Blockchain}\label{BlockLayers}
Since blockchain is a fully functional P2P network having the features of decentralized communication, incentives, and consensus, thus, to understand the functionality of this a bit further, researchers divided it into multiple layers. Various works have been carried out to identify and discuss different layers, for instance, Homoliak~\textit{et al.} in~\cite{survey10} and Wu~\textit{et al.} in~\cite{survey06} classified blockchain into four layers named as network layer, consensus layer, data/state layer, and application layer. \rev{Similarly, Belotti~\textit{et al.}~\cite{survey07} divided the architecture of blockchain into five layers and added another layer named `Execution layer’.} Similar to this, Xie~\textit{et al.} in~\cite{introanomaly05} added another layer and proposed a 06 layered architecture of blockchain comprising of data layer, network layer, consensus layer, incentive layer, contract layer, and application.\rev{Since, the focus of our work is to detect anomalies in the blockchain network, thus, after going through the available literature, it can be concluded that four layers of blockchain are more prone to anomaly attacks. Therefore, in this article, we discuss the detection of anomalies in blockchain from the perspective of four layers named as data layer, network layer, incentive layer, and contract layer.} \\
\rev{Each layer of blockchain has its own functionalities and has corresponding tasks associated with it. In this section, we briefly highlight the functionalities of four prominent layers which we will be discussing later \rev{in this article from an anomaly detection perspective.} However, readers interested in understanding the functioning of blockchain layers from the perspective of their functioning, applicability, and deployment can study the discussion provided in~\cite{revsurvey01,revsurvey02, survey10, survey06, revsurvey03, revsurvey04}.}

\subsubsection{Data Layer}
\rev{The first layer in the blockchain architecture is data layer, which comprises of data blocks. The blocks in the data layer are time-stamped and are linked with one another via hashes to form a chain-like structure. A typical block in a blockchain network \final{comprises of two} parts named as header and body of block~\cite{introanomaly05}. Block header mainly comprises important metadata parameters,} such as block hash, previous block hash, time-stamp, nonce value, Merkle root hash, etc. The contents in the block header can also differ according to the need of application. \rev{The second part of the block is the block body which mainly comprises the transactions which are picked from the mining pool for the purpose of storing on blockchain. The hashes of these transactions are computed and are further combined to form a single Merkle root hash, which is also the part of the block header.}

\subsubsection{Network Layer}
\rev{Network layer in blockchain is responsible for carrying out distributed communication and networking for blockchain peers. Since, blockchain is a P2P network in which all peers have the same rights,} thus, the functionality of this layer is to run such networking and communication models which ensure the timely distribution, forwarding, and verification of blocks in the network~\cite{survey10}. For instance, if a transaction is generated in the blockchain network, then the network layer is responsible to broadcast this transaction to all neighbouring peers. Similarly, verification acknowledgement of this transaction will also be returned \rev{via using functionalities of the network layer.} Afterwards, if the transaction turns out to be a valid transaction, then it will again be sent to broadcast to other peer nodes. Contrarily, an invalid transaction is denied and is not sent for further broadcast in the network. A detailed discussion about transaction signing and verification is out of scope of this article, interested readers are suggested to study the discussion given in~\cite{survey07}.

\subsubsection{Incentive Layer}

Incentive layer in blockchain revolves around financial incentives in the network, which serves as a major factor of motivation for participants of the network~\cite{survey07}. In a decentralized network with no centralized authority, maintaining motivation of participants is a major challenge, and this challenge in \rev{blockchain is solved by developing incentive \final{models through} the incentive layer. For instance, in the Bitcoin network, a specific number of Bitcoins are given as a reward upon completion of a round of mining. This reward mechanism motivates Bitcoin users to actively participate in the mining process.} Similarly, in other blockchain networks, similar rewards are issued upon completion of specific tasks, which serves as a driving force for the network. Apart from incentives, certain penalties and deposits do also come under the scope of this layer. 

\subsubsection{Contract Layer}

Contract layer \final{(also known as smart contract layer)}, is responsible to bring programmable functionalities in the blockchain network. Modern blockchain networks, such as Hyperledger Fabric, Ethereum 2.0, etc., provide the functionalities of dynamic programming in which the users can write a logical program to execute it on the network. The program is written in the form of a contract, which is known as smart contract. This smart contract is a piece of executable code, which runs over the blockchain network, and performs the tasks assigned to it. There could be multiple types of smart contracts depending upon the nature of execution, application, and requirements. \final{A detailed} discussion about different types of smart contracts is out of scope of this article, interested readers are suggested to study an interesting article by~\cite{preanomaly01}. 

\begin{figure*}
\centering
  \includegraphics[scale=0.4]{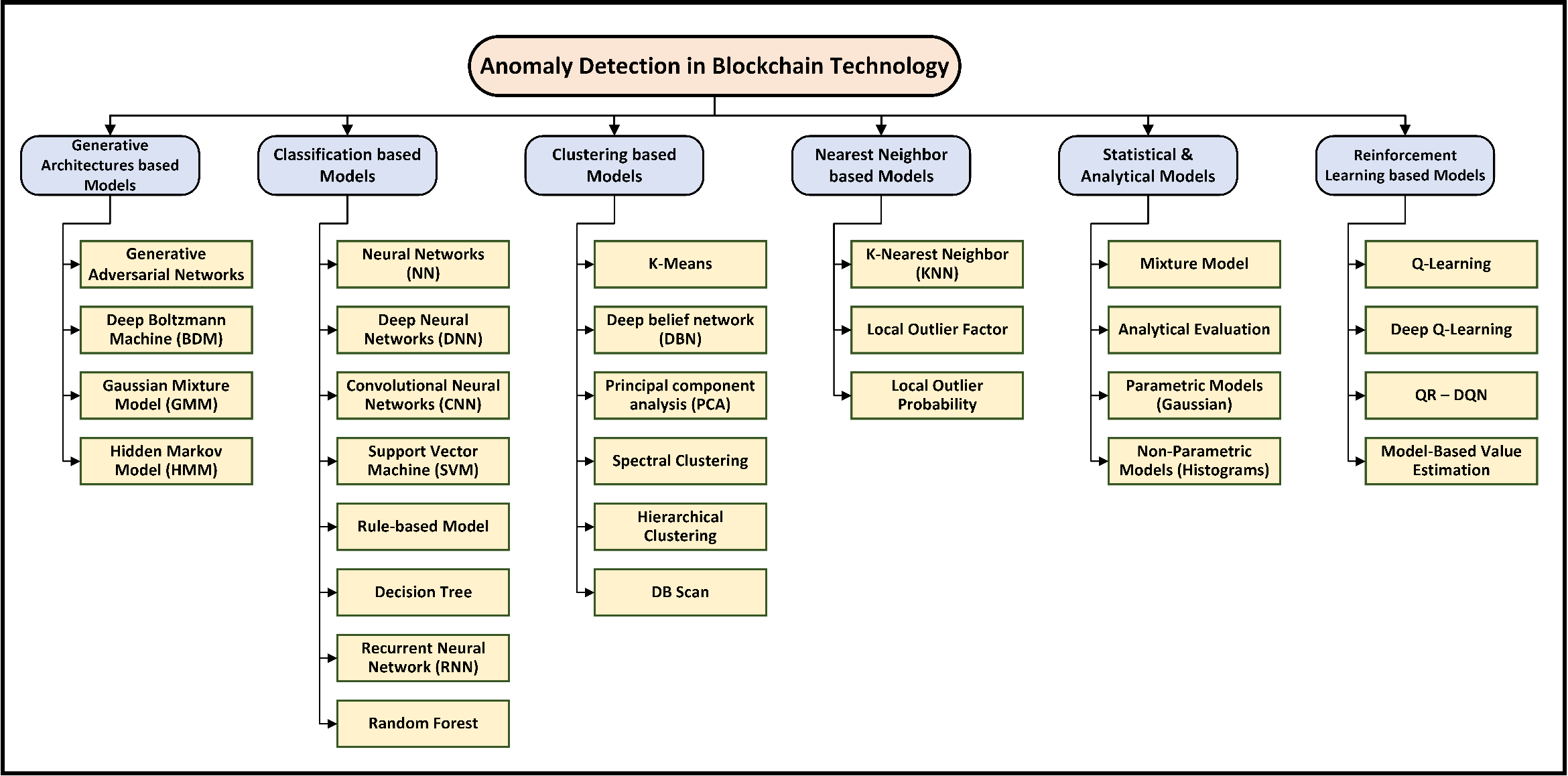}
  \caption{\mubcom{Classification of Anomaly Detection Models in Blockchain Technology}}
  \label{fig:anomalyclassification}
\end{figure*}

\subsection{\rev{What is Anomaly Detection?}}
\rev{In this section, we provide a brief discussion about anomaly detection for a generic audience.} %

\subsubsection{\rev{What is Anomaly?}}
\rev{Anomaly can be termed as a specific pattern in the collected or transmitted data which does not show a well-suited regular behaviour~\cite{comstref02}.} Anomalies can be caused due to multiple reasons, ranging from human and non-human errors to maliciously controlled activities. Anomalies are also known as novelties, noise, outliers, malicious activities, and exceptions in various perspectives. However, it is important to note that anomalies can be both harmful or harmless depending upon their nature and origin. For instance, an anomalous transaction which can suddenly reduce or increase the price of a selective share to the verge of market crash is considered a harmful anomaly. Contrarily, a sudden transaction of a high number of shares from a specific node, which will not have any significant impact over the market will be classified as a harmless anomaly. No matter harmful or harmless, it is important to detect the presence of anomalies in a network within an actionable time in order to reduce the risk of \final{any future catastrophe.}

\subsubsection{\rev{What is Anomaly Detection System?}}
\rev{Detection \final{of anomalous} behaviours in the network is termed as anomaly detection, which is carried out by designing and implementing sophisticated anomaly detection models. Anomaly detection models play a very significant role in order to detect the anomalies in an efficient manner. In order to detect anomalies within an actionable timeframe, a continuous monitoring of network behaviour is required. In order to do so, an anomaly detection system is \final{usually designed,} which continuously monitors the changes in the \rev{network with the help of the most appropriate detection model in accordance} with the network requirement. Usually, monitoring in these anomaly detection systems is categorized into two subtypes named as passive monitoring and active monitoring. There are multiple definitions of active and passive monitoring, however, generally speaking, in active monitoring, data is usually generated and monitored on a specific part of the network in order to identify the anomaly in the selected region. Contrarily, passive monitoring is referred to as monitoring of the performance of the whole network in order to identify any anomalous behaviour. Despite the difference in functioning, the end goal of both types of monitoring systems is to identify the anomalous properties of the network and categorize whether they are harmful or harmless.}

\subsubsection{\rev{Selecting Optimal Anomaly Detection Model}}

\rev{The actual question in anomaly detection research is ‘which is the best anomaly detection model?’. As per the discussion in various research works, it will not be wrong to say that there is no single unified algorithm which will work best in all scenarios, which means that the optimal model can vary according to the type and requirement of the problem scenario. Similarly, each of the models has its associated advantages and disadvantages. For example, if one already has a labelled set of records, then supervised machine learning models will outperform others. Similarly, if we have data with a small number of labels, then semi-supervised models are the best options. Contrarily, if we have a completely novel input data with no labels, then unsupervised learning models outperforms other models. Alongside data, it is also important to identify the type of anomaly we are looking for. For instance, if one is looking for small variations of patterns in the collected \final{data which} occur in a non-systematic way, then outlier detection models perform better, such as local outlier factor, etc. Similarly, if one is trying to identify an abrupt or systematic change in the data in comparison with normal behaviour, then classification and clustering based models turn out to be the optimal choice. However, if one wants to analyse slow and long-term modifications and changes in the network, then certain statistical and analytical models can be used to carry out analysis. Therefore, choosing the optimal model is mostly dependent upon the input data, tackled anomaly, and desired outcome. In the later sections, we highlight the technical challenges and evaluation \final{metrics that} play a significant role in determining the optimal anomaly detection for blockchain based application. A detailed classification of multiple anomaly detection models and their functioning from the perspective of blockchain has been provided in Section~\ref{AnomalyBlockModels}.}

\subsection{Anomaly Problem in Blockchain Layers}
\rev{To efficiently run operations of a blockchain network, it is important to identify and take action against adversarial behaviours within the actionable time frame. In order to do so, anomaly detection models came into discussion, which  are responsible for effective detection of anomalous behaviour of a specific node. Generically, we divide the anomaly detection models into six sub-categories on the basis of their functioning, which are named as generative architectures, classification based models, clustering based models, nearest neighbour models, statistical \& analytical models, and reinforcement learning based models (see Fig.~\ref{fig:anomalyclassification}).}\\
\rev{From the perspective of blockchain layers, it is important to mention that the anomalies in blockchain technology are not pretty generic because almost all layers of blockchain have their specific anomalies, thus, their detection \final{mechanism vary}. For example, an anomaly related to a Ponzi scheme falls under the category of incentive layers, while spreading of anomalous messages in the network falls under the category of network layer. Therefore, it is important to classify these anomalies according to the layer they fall under in order to carry out their efficient detection.}\\

\subsection{Technical Challenges while Integration of Anomaly Detection in Blockchain}

Nevertheless, development of basic anomaly detection models is not much complicated if one has basic knowledge of machine learning. However, when it comes to anomaly detection in blockchain networks, then certain challenges arise due to the nature of blockchain. \rev{In this section, we discuss certain prospective challenges that one can face while developing anomaly detection models for blockchain based scenarios.}

\subsubsection{Network-wide Consensus on Anomaly}
The first challenge that one has to overcome while developing anomaly detection models for blockchain networks is to carry out a network-wide consensus on anomaly. Since, the blockchain network has no centralized entity to determine rules, thus, it becomes hard to categorize a specific event to be an anomaly or not. Therefore, alongside designing a mechanism to detect outliers, one also has to make sure that this outlier is considered as an anomaly throughout the network. \rev{In short, complete \final{network has} to reach a consensus that a particular event is an anomaly and appropriate actions should be taken against it.} This becomes even more challenging, when some nodes in the network start behaving in a malicious manner. Therefore, considering the aspect of network-wide consensus alongside designing anomaly detection models is important in blockchain because of decentralization.
\subsubsection{Careful Selection of Outlier Features}
Since blockchain is a novel paradigm and plenty of attacks are pretty new even for researchers, therefore, selecting the best features for outliers is one of the major challenges. For instance, if one wants to label the purpose of a smart contract in a blockchain network, then it becomes hard due to very less available references. \final{To overcome} this, researchers worked over developing automated methods for labelling new and unknown smart contracts for detection of possible anomalies from these labels~\cite{preanomaly02}. \rev{However, the majority of smart contracts are pretty identical and do not show significant differences, therefore, such methods are not well-established till now. Thus, it is always challenging to identify these features and one has to be extra careful} in order to ensure that they do not categorize a legitimate user or transaction as anomaly. 

\subsubsection{Smart Contract Programmability \& Execution due to Environmental Constraints}
It is important to mention that majority of smart contracts of blockchain are being developed in bytecode rather than binary code, which makes it difficult to run traditional anomaly detection models on blockchain network in real-time environment~\cite{contract07}. Similarly, detecting anomalies at bytecode level becomes even more challenging because not all information is available at the level of bytecode, and certain information gets lost during the compilation process~\cite{contract08}. Therefore, designing such models, which efficiently detect anomalies from smart contract bytecode is a big challenge that every researcher working in blockchain anomaly detection faces regularly.

\subsubsection{Lack of Defined Rules}
\rev{With the advent of every new application of blockchain, the rules change accordingly. E.g., a blockchain based smart grid will have a different set of rules for anomaly as compared to a network of blockchain based electric vehicles. Therefore, the rules defined for anomaly detection in a decentralized smart grid cannot be applied to other blockchain networks.} Similarly, the generic rules defined for generalized blockchain networks cannot be applied to specific domain-oriented networks. Certain researchers worked over carrying out manual inspection of models and truncations in order to gather detailed information, however, it is a tiresome and inefficient process~\cite{contract07}. Therefore, it can be said that designing a set of rules is one of the major challenges for researchers working in the domain of anomaly detection in blockchain networks.

\section{Fundamentals Guidelines to Identify Anomalous Behaviour in Blockchain}

\begin{figure*}
\centering
  \includegraphics[scale=0.55]{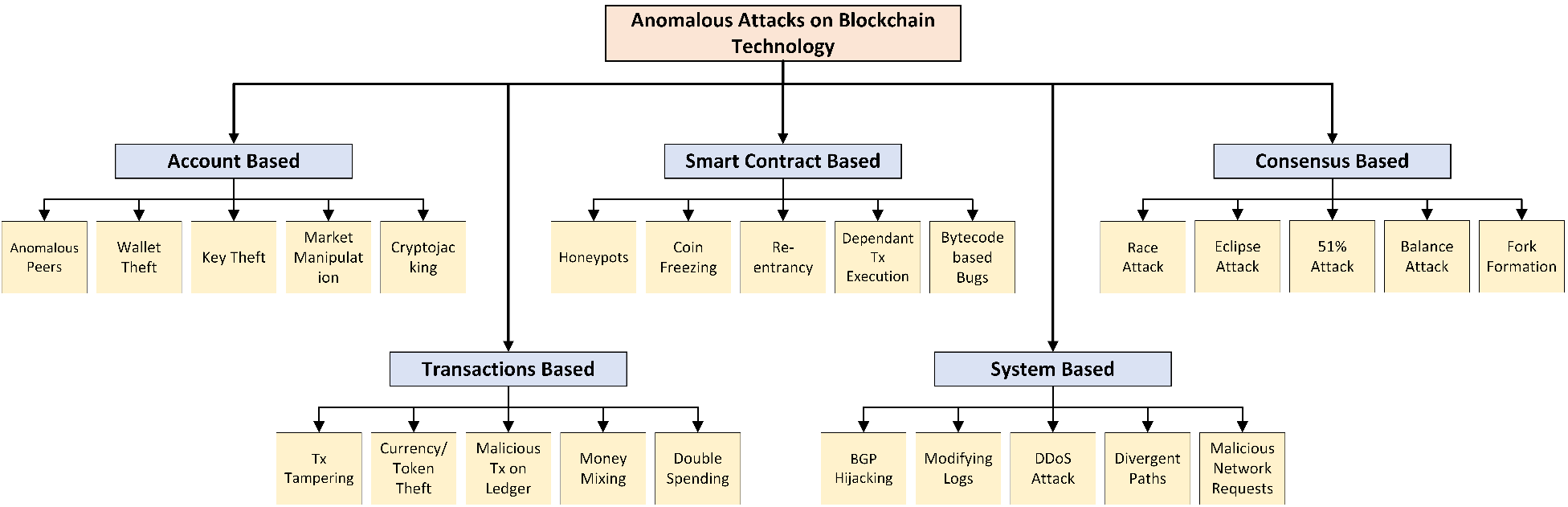}
  \caption{\rev{Classification of Anomalous Attacks on Blockchain Technology}}
  \label{fig:anomalyattacks}
\end{figure*}

\subsection{Types of Anomalous Attacks on Blockchain}

\subsubsection{Malicious Transaction Pattern Detection}
One of the most common anomalies in blockchain networks is uneven transactions. Due to the pseudonym property of blockchain, nodes usually feel safe to carry out large transactions, however, among these transactions, some transactions are also uneven, \rev{which are mostly carried out for some malicious purposes. For example, some users try to carry out money laundering} while being anonymous in the blockchain network~\cite{fundanomaly01}. \rev{Therefore, it is important to identify such transactions before recording them to the ledger in order to take appropriate action against them. Fortunately, due to the decentralized nature of blockchain, these transactions} can be identified by observing various transaction patterns.
\subsubsection{Double Spending Detection}
As the name suggests, double spending is related to spending or utilization of an asset more than once in a decentralized network. Since blockchain is decentralized and there is no central authority to verify each transaction, therefore, malicious nodes usually try to use this nature of blockchain to carry out double-spending~\cite{fundanomaly02}. A transaction on blockchain is finalized once it gets validated by the peer nodes, however, during the process of validation, some malicious nodes try to use the same amount of funds to carry out multiple transactions. Due to strong consensus guarantees of blockchain networks, it is not that easy to carry out double spending, however, sometimes hackers succeed in fooling the network. Therefore, for successful functioning of a blockchain network, it is important to timely detect and even predict the occurrence of double spending in the network.

\subsubsection{\rev{Market Manipulation}}
\rev{Manipulation in blockchain is not just limited to cryptocurrencies and it can be done in any sort \final{of asset which is} associated with a financial factor. Moreover, a number of manipulation strategies have been identified by researchers and experts, however, generally, the three most famous ones are pump and dump, wash trading, and whale wall spoofing. In pump and dump, an individual or a group of individuals try to artificially increase the price/value of an asset or a cryptocurrency by developing a fake sense of attention towards it~\cite{comstref18}. Another market manipulation strategy is wash trading, in which a group of individuals usually try to buy and sell a specific cryptocurrency/asset rapidly in order to create a sense of an interest in the market~\cite{comstref19}. Sometimes, this strategy is also adopted by exchanges which are operating at a small  scale and are unregulated. The third strategy named as whale wall spoofing is pretty similar to wash trading, in which a big whale (usually an entity with a huge asset) tries to create buying or selling orders in the market for the corresponding asset/coin. This is usually done in order to trick the relevant investors/users into panic buying and selling. Considering the intensity of market manipulation, it is important to determine the occurrence of \final{such event within} an actionable time. Certain research works have been carried out in this regard which developed machine learning and statistical analysis based market manipulation models, however, there is still a lot of room for improvement in this domain.}

\subsubsection{Money Mixing Detection}
Generally, money mixing is a legitimate process in blockchain networks, which revolves around mixing different assets or tokens to overcome identifiability trail~\cite{fundanomaly03}. The mixing in blockchain is usually used to enhance transaction anonymity from malicious attackers. However, some maleficent nodes try to take unlawful advantages out of it and try to hide their transaction patterns to carry out immoral activities, such as money laundering, etc. \rev{Therefore, it is important to identify malicious money mixing in order to protect the blockchain network from unlawful activities.}

\subsubsection{Currency/Token Theft Detection From Network}
Apart from basic anomalies, things get intense when malicious users try to steal tokens directly from users by hacking or similar other ways. \rev{This is one of the most common attack types and multiple incidents have been highlighted in the past where hackers stole millions of dollars from investors at different occasions. For example, it has been reported that in 2020 \final{alone that blockchain} hackers stole an approximate sum of \$3.8 Billion in 122 different attacks over the network~\cite{comstref20}.} Contrary to centralized payment systems (such as banks), blockchain network has no centralized mediator to regulate transactions. \rev{Therefore, the importance of theft detection in decentralized blockchain scenarios increases exponentially to prevent any large mishap.}

\subsubsection{Smart Contract Anomalies}
Smart contracts play a critical role in functioning and development of modern day blockchain because they add the feature of programmability in blockchain networks. Through this programmability feature, one can use blockchain for numerous advantages ranging from tracking decentralized ownership of assets to verification of education degrees. However, the base of these smart contracts is programming, and programming is not guaranteed to be 100\% perfect all the time. There is always a possibility of mistakes in the smart contract, and since smart contracts are irreversible, thus, these mistakes can cause big damage. Similarly, apart from unintentional mistakes, some adversaries try to set up honeypots in smart contracts, the sole purpose of which is to perform fraudulent activities such as coin theft, etc~\cite{contract08}. Therefore, in both of these cases, it is important to identify any anomaly in blockchain smart contracts before its execution.

\subsubsection{Malicious Network Requests}
From an \final{outsiders perspective,} blockchain is a secure network, however, from an insider point of view, blockchain network is still prone to certain attacks, and malicious network requests are one of them. In such requests, adversary nodes try to tamper with the transactional values before pushing these values to peer nodes. In this way, hackers try to divide the network into multiple parts so that they would not be able to communicate with each other. These types of malicious requests are also known as routing attacks on the network, and they can further be divided into partition and delay based attacks on the basis of their nature. These malicious requests can cause a big harm in the network; thus, their timely identification and eradication is compulsory.

\subsubsection{Divergent Path \& Forks}
Blockchains as a ledger are immutable according to their nature, which implies that the information on the ledger cannot be changed once it gets recorded. However, if an adversary \final{tries to play} maliciously while trading or while developing a smart contract, then they can exploit this feature of immutability for their unlawful benefits and can initiate formation of new forks in the network. \rev{One of the prime examples of fork formation are the forks originated out of the Bitcoin network, such as Bitcoin Cash, etc~\cite{comstref22, comstref23}.} Attackers usually try to create divergent paths in order to take control over 51\% of the network, which basically leads to disastrous consequences. Therefore, for successful functioning of a blockchain network, the timely detection and prevention of forks is very important.

\subsubsection{Tampering Blockchain Logs}
\rev{In certain blockchain based applications, logs play an important role in determining operational activity. For instance, in a manufacturing industry, whenever a new step is performed, it is recorded into a blockchain ledger by log system,} which is then used to ensure the quality of product. However, if some nodes try to act as an adversary, then they can try to tamper with the logs in order to misguide the scrutinizing body, which makes audit difficult or sometimes \final{even impossible.} Therefore, in order to ensure successful functioning of such blockchain applications, a regular and thorough analysis of logs is required to overcome any mishap.

Furthermore, a detailed taxonomy of these anomalies can be visualized in Fig.~\ref{fig:anomalyattacks}.
 
\subsection{\rev{Categorization and Functioning of Anomaly Detection Models for Blockchain Technology}}\label{AnomalyBlockModels}

\rev{From the viewpoint of functioning, anomaly detection models can be categorized into multiple types ranging from machine learning based to statistical anomaly detection models. A graphical illustration showing the detailed taxonomy of anomaly detection models has been provided in Fig.~\ref{fig:anomalyclassification}. The first type of anomaly detection models are generative architecture based models, which can be defined as the use neural network based algorithmic architectural methods to generate new synthetic data instances, which is further used to train the anomaly detection models accordingly in a manner that when one puts the anomalous input data, the model should be able to highlight the reconstructed anomalous data on the basis of reconstruction error~\cite{comstref03}. These models are helpful in categorization of anomalies, outliers, and novelties in the data, especially transactional data of blockchain. Certain popular models in this category are generative adversarial networks (GANs), deep Boltzmann machine (DBM), Gaussian mixture model (GMM), and hidden Markov model (HMM), etc. The second type of anomaly detection models are known as classification based anomaly detection models~\cite{comstref04}. These are usually typical machine learning based models with an aim to classify the data into two subtypes (normal and anomalous) on the basis of detection pattern. In blockchain, these models can be used to identify the anomalous behavior of any peer. A large number of machine learning based algorithms can be categorized in this domain, although, some of the prominent ones are neural networks (NN), deep neural networks (DNN), convolutional neural networks (CNN), support vector machine (SVM), etc.} \\
\rev{The next category is clustering based anomaly detection models, which works over the phenomenon of making clusters of the training data and then predicting the new outcome with relevance to the developed cluster~\cite{comstref05}. For instance, in DB Scan (a famous clustering based anomaly detection model) develops density based clusters of the data during the training phase, and then during testing phase, it compares the new received data in accordance with these already developed clusters in order to classify the normal or anomalous behaviour. These cluster based analyses have been proved to be very effective in analyzing transaction frequency and degree in blockchain transactions. Some of the renowned models in this category are k-means clustering, deep belief network (DBA), DB Scan, etc. Another category very similar to clustering based approach is nearest neighbour models, which basically classify the training data into neighbours on the basis of distance parameters~\cite{comstref08}. These models work over the belief that the normal blockchain activities will occur in the dense areas where multiple nodes are neighbours of each other. However, anomalous activities occur far from their nearest neighbouring nodes. Some famous models in this category are k-nearest neighbours (KNN), local outlier factor, local probability outlier, etc.} \\
\rev{The next category is statistical and analytical analysis based blockchain anomaly detection~\cite{comstref07}. These are the simplest and traditional anomaly detection models as compared to all other categories as they do not require any complex algorithmic processing to identify anomalies in the data or network. Because of their simple structure they are usually used in scenarios which have less computational capability. Despite being simple, they are quite powerful in identifying various baseline anomalies in the blockchain network. Some of the prominent statistical and analytical models are parametric models (such as Gaussian), and non-parametric models (such as Histogram, Boxplot). These models are usually used to highlight anomalous transactions, which are usually linked with anomalous gas usage for Ethereum blockchain. Similarly, these models have also proved their effectiveness in analysis of bugs in online/offline smart contracts, such \final{as honey-pots.} The last category is reinforcement learning based anomaly detection models, which basically works over the phenomenon of rewarding the training model upon identification of blockchain anomalies in the correct manner~\cite{newref07}. Some of the renowned reinforcement learning models that can be used to carry out anomaly detection are Q-learning, deep q-learning, QR-DQN, and model based value estimation.}

\subsection{Evaluation \final{Metrics} being used to Identify Anomalies}
In an anomaly detection system, one needs to be very precise about every factor they consider, e.g., one cannot overlook certain anomalies as it can lead to serious mishap. Similarly, on the other hand one cannot even classify a normal behaviour as an anomaly, which can also lead to strenuous trouble certain times. Therefore, while developing such models, researchers are required to carefully consider certain factors, such as accuracy, precision, etc. In this section, we provide a brief overview of such factors alongside their importance in anomaly detection.
\subsubsection{Sensitivity \& Specificity of Outcome}
In anomaly detection, the factors related to sensitivity and specificity play a critical role in determining effectiveness of any model. Nevertheless, in an anomaly detection model, \rev{the decision could be right or wrong which means true or false} but quantifying the outcome to make the model more effective is the key. In order to understand the sensitivity and specificity of an anomaly detection model a bit further, researchers have devised certain terms, which are discussed as follows:
\paragraph{True Positive (TP)} The outcome TP means that the anomaly model gave a positive outcome and identified a specific behaviour as an anomaly, and in reality the result is true and that behaviour was actually anomalous.
\paragraph{False Positive (FP)} This term means that the anomaly model gave a positive outcome and identified the node/behaviour as an anomaly, but in reality it was not anomaly.
\paragraph{True Negative (TN)} The outcome TN identifies that the anomaly detection model gave a negative result for detection of anomalous behaviour and in reality its true and the behaviour was not anomalous.
\paragraph{False Negative (FN)} The outcome FN means that the anomaly model gave a negative detection of anomaly but in reality the behaviour was anomalous.\\
In an ideal anomaly detection model, the rate of TP and TN should be high, and the rate of FP and FN should be low.
\subsubsection{Confusion Matrix \& Accuracy}
Confusion matrix (also known as error matrix) is a visual \final{table which} is used by researchers to analyse \rev{the efficiency and accuracy of any model.} In anomaly detection mechanisms, it is used to carry out comparison between the predicted and actual class labels. E.g., in a 2-by-2 confusion matrix, one side would represent the actual/true values, \rev{and the other side would represent predicted values. The matrix is then filled according to the outcomes of the model in comparison with the actual values. E.g., if the predicted value is Yes, and the actual outcome is also Yes, then the value of TP in the table is incremented. Similarly, if the value of prediction \final{is `No',} and the actual value is also No, then the value of TN is incremented.} In this way, all values of TP, TN, FP, and FN are filled, and are then used to compare accuracy of model via following equation~\cite{fundanomaly06}:
\begin{equation}\label{equationCM}
Accuracy = \frac{TP + TN}{TP +FP + FN +TN}
\end{equation}

\subsubsection{Recall, Precision, and F-Score}
Recall, precision, and F-score are \final{the usual factors} which are being used by researchers to analyse outcome of an anomaly detection model.  The first terminology among them \rev{\final{is recall which} can be defined as the number of correctly identified instances by a model.} To be more detailed, it is the number of true positives divided by the total number of actual positive instances. Precision can be termed as a counterpart of recall, as the precision is the ratio between total number of correct returned results and cumulative sum of positively identified results including false positives. \rev{F-Score, also known as F-1 Score, which mixes the property of precision and recall in a harmonic manner, so that the model can be evaluated in the best manner. The higher F-Score means the credibility of the model to give good results is high.} A detailed discussion about these parameters is out of scope of this article, readers interested in studying these parameters can study the interesting article by Bhuyan~\textit{et al.}~\cite{fundanomaly06}.


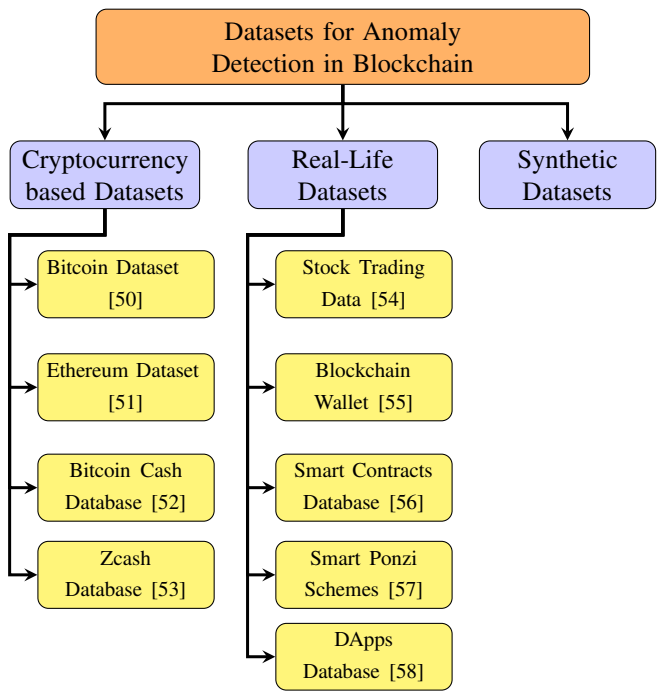
\begin{figure}[]
     \centering

\begin{tikzpicture}


\node [firstblock,  text centered,  text width=18em] (a1) {Datasets for Anomaly Detection in Blockchain};


\node[secondblock, below of=a1, yshift=-2em, xshift=-9em, text width=6.5em](b1){Cryptocurrency based Datasets};
\node[secondblock, below of=a1, yshift=-2em, xshift = 0em, text width=6.5em ](b2){Real-Life Datasets};
\node[secondblock, below of=a1, yshift=-2em, xshift = 8.5em, text width=6em](b3){Synthetic Datasets};


\node[fourthblock, below of=b1, xshift = 0.8em, yshift=-1.3em, text width = 6em] (b1c1) {\footnotesize{Bitcoin Dataset\newline~\cite{dataset01}}};
\node[fourthblock, below of=b1, xshift = 0.8em, yshift=-5.2em, text width = 6em] (b1c2) {\footnotesize{Ethereum Dataset\newline ~\cite{dataset02}}};
\node[fourthblock, below of=b1, xshift = 0.8em, yshift=-9.0em, text width = 6em] (b1c3) {\footnotesize{Bitcoin Cash Database~\cite{dataset03}}};
\node[fourthblock, below of=b1, xshift = 0.8em, yshift=-12.3em, text width = 6em] (b1c4) {\footnotesize{Zcash Database~\cite{dataset04}}};

\node[fourthblock, below of=b2, xshift = 0.8em, yshift=-1.3em, text width = 6em] (b2c5) {\footnotesize{Stock Trading Data\cite{dataset05}}};
\node[fourthblock, below of=b2, xshift = 0.8em, yshift=-5.2em, text width = 6em] (b2c1) {\footnotesize{Blockchain Wallet~\cite{dataset08}}};
\node[fourthblock, below of=b2, xshift = 0.8em, yshift=-9.0em, text width = 6em] (b2c2){\footnotesize{Smart Contracts Database~\cite{dataset06}}};
\node[fourthblock, below of=b2, xshift = 0.8em, yshift=-12.3em, text width = 6em] (b2c3) {\footnotesize{Smart Ponzi Schemes~\cite{dataset09}}};
\node[fourthblock, below of=b2, xshift = 0.8em, yshift=-15.4em, text width = 6em] (b2c4) {\footnotesize{DApps Database~\cite{dataset07}}};


\path [line] (a1)-- ($(a1.south)+(0,-0.25)$) -|(b1);
\path [line] (a1)-- ($(a1.south)+(0,-0.25)$) -|(b2);
\path [line] (a1)-- ($(a1.south)+(0,-0.25)$) -|(b3);


\path [line] (b1.south)|- ($(b1.west)+(0,-0.8)$) |-(b1c1);
\path [line] (b1.south)|- ($(b1.west)+(0,-0.8)$) |-(b1c2);
\path [line] (b1.south)|- ($(b1.west)+(0,-0.8)$) |-(b1c3);
\path [line] (b1.south)|- ($(b1.west)+(0,-0.8)$) |-(b1c4);

\path [line] (b2.south)|- ($(b2.west)+(0,-0.8)$) |-(b2c1);
\path [line] (b2.south)|- ($(b2.west)+(0,-0.8)$) |-(b2c2);
\path [line] (b2.south)|- ($(b2.west)+(0,-0.8)$) |-(b2c3);
\path [line] (b2.south)|- ($(b2.west)+(0,-0.8)$) |-(b2c4);
\path [line] (b2.south)|- ($(b2.west)+(0,-0.8)$) |-(b2c5);

\end{tikzpicture}
	\small \caption{Datasets for Anomaly Detection in Blockchain}
     \label{fig:dataset}
\end{figure}


\subsection{Key Requirements for Anomaly Detection Mechanisms Design}
Usually, behaviour of an anomalous participant varies noticeably as compared to other legitimate participants, \rev{therefore, they get identified by anomaly detection models. However, in order to get efficient results, certain key requirements need to be considered while designing anomaly detection models for blockchain networks.} In this section, we discuss these requirements to give readers an overall viewpoint.

\subsubsection{Data Collection Requirement}
\rev{One of the major requirements for efficient functioning of any anomaly detection model is to have an adequate amount of data for analysis. Even this step of data collection is challenging in the traditional anomaly detection environment, and in blockchain based anomaly detection scenarios, it becomes even more challenging due to various protecting mechanisms in the way of blockchain data collection.} For each type of blockchain, the data collection methods do vary, e.g., in case of public blockchain, the data is available to all participating nodes and one can carry out anomaly detection easily. However, in case of private or consortium setting, the data is not publicly available, and certain approvals are required before carrying out any processing over data. Furthermore, in all these types, the participating nodes are usually identified by pseudonyms, therefore, tracking the exact individual even after classifying it as an anomaly is sometimes very complicated due to lack of data about that individual. \rev{Alongside these issues, it is equally important to protect the privacy of participating blockchain users/individuals in order to ensure the trust in the network. Therefore, in certain cases one also has to integrate certain privacy preservation mechanism as well with blockchain technology in order to enhance privacy.}\mubcom{Some of the key datasets, which can be used to train models for future prediction of anomalies of blockchain network have been highlighted in Fig.~\ref{fig:dataset}.}

\subsubsection{\mubcom{Data Preprocessing Requirements}}
\mubcom{Raw data usually contains a lot of noise, therefore, preprocessing is a step in anomaly detection, in which collected data is modified and manipulated in order to reduce any noise and vulnerabilities from the data which can cause hurdle in detection of anomalies~\cite{newref01}. Some of the key steps involved in data preprocessing include cleaning of data, transformation of data, selecting required features from data, reduction of data, and discretization of data. These steps are there to ensure that only the fine-grained data is fed to anomaly detection models, so that anomaly is detected as quick as possible. \rev{In the majority of anomaly detection models, data preprocessing is considered as an essential step before feeding any data to anomaly detection models in order to enhance the detection accuracy and efficiency.}} \rev{A large number of challenges are associated with data preprocessing, however, some of the prominent ones that blockchain also faces is inconsistency of data, missing data, and random pseudonyms, etc.}\\
\mubcom{Similar pathway is also adopted in blockchain based anomaly detection \final{models in which} a data collected via blockchain is nodes is preprocessed via pre-developed mechanisms, usually via preprocessing smart contracts~\cite{newref02, contract04}. Similarly, in certain bytecode based anomaly detection models of blockchain, data is denoised using autoencoders~\cite{contract07}. Comparably in detection of anomalies in HYIP on blockchain, the preprocessing phase usually \rev{involves removal of the transaction change part alongside calculation and identification of patterns} between transactions~\cite{data07}. From the perspective of malicious account detection on Ethereum, the data is usually preprocessed in two steps, first via string comparison to identify duplicate addressed and then via filtering of EOA addressed and contract addresses~\cite{data13}. Apart from the traditional preprocessing, certain real-time big data preprocessing tools and methods have also been developed till now for various applications, which modifies the steps involved in preprocessing in a way that it enhances the overall time and efficiency of preprocessing~\cite{newref03}. However, such works have not yet been carried out in the field of blockchain technology and there is a need for such integrations in future. }

\subsubsection{Computational Requirements}
In order to detect anomalous behaviour in an efficient manner, the model needs to predict the anomalies within a specific time-frame, and that can only happen if the computational requirement to run an anomaly model matches with the required task. Certain models require high computational complexity, while on the other hand some models only need minimal \final{computational power.} \rev{Similar is the case with blockchain, as for certain types of anomalies, which require processing of large amounts of data, such as smart contract bugs detection or malicious transaction identification, one needs to learn from a large amount of data, which require large computational power. Contrarily, for some basic graph/cluster based analysis over data, \rev{the requirement of computational power is pretty less comparatively.} In blockchain anomaly detection, a large computational power is usually consumed during the mining and consensus process. Therefore, in order to avoid overloading of machines, the on-chain anomaly detection models need to require computational power alongside high accuracy for smooth functioning of the network.} Various generic models have been designed to reduce computational cost of anomaly detection, however, in blockchain, this field is still progressing, and it has a lot of room in it for research.

\subsubsection{Accuracy Requirements}
Anomaly detection accuracy is another key element that cannot be ignored while developing anomaly detection models. This element even strengthens in case of  blockchain anomaly because in blockchain decisions are irreversible and will always be there on the ledger. E.g., if anomaly detection predicts some specific node as an anomaly, and authorities take some action just on the basis of that anomaly detection outcome, then this action will remain on blockchain forever. This is fine in case of an anomalous node, but in case of a false positive result, it will be a big challenge for authorities as they have identified a rational user as an anomaly, which can lead to disastrous outcomes. \rev{As in case if a penalty smart contract is executed on an honest node, then it will reduce the trust in the network drastically. Therefore, it is very important for a blockchain anomaly detection model to have high accuracy during prediction.}

\section{Anomaly Detection in Data Layer of Blockchain} \label{DataLayerSection}

\begin{figure*}[]
     \centering

\begin{tikzpicture}


\node [firstblock,  text centered, minimum width = 25em,  text width=25em] (a1) {Anomaly detection in Data Layer \\ Sec.~\ref{DataLayerSection}};


\node[secondblock, below of=a1, yshift=-3em, xshift=-20em, text width=6.5em](b1){Bitcoin Fraud Detection};
\node[secondblock, below of=a1, yshift=-3em, xshift = -8.8em, text width=6.5em ](b2){Framework Design for Anomaly};
\node[secondblock, below of=a1, yshift=-3em, xshift = 0.2em, text width=6em](b3){Behaviour Pattern Classification};
\node[secondblock, below of=a1, yshift=-3em, xshift = 9.8em, text width=5em](b4){Ethereum Fraud Detection};
\node[secondblock, below of=a1, yshift=-3em, text width=6.5em, xshift = 19.1em](b5){Tx Relative Similarity Clustering};



\node [fourthblock, below of=b1, xshift = -4.5em, yshift=-5.5em, text width = 3.3em] (b1c1d1) {\footnotesize{Tx \& User Deanony- mization\newline ~\cite{data01, data09}}};
\node [fourthblock, below of=b1, xshift = 0em, yshift=-5.5em, text width = 3.3em] (b1c1d2) {\footnotesize{Integrating Autoencodersfor Anomaly ~\cite{data08, newref08}}};
\node [fourthblock, below of=b1, xshift = 4.5em, yshift=-5.5em, text width = 3.3em] (b1c2d1) {\footnotesize{Tx Frequency Analysis\newline ~\cite{data05, data11}}};

\node [fourthblock, below of=b2, xshift = -2.3em, yshift=-5.5em, text width = 3.32em] (b2c1d1) {\footnotesize{Blockchain Health Analysis~\cite{data06}}};
\node [fourthblock, below of=b2, xshift = 2.3em, yshift=-5.5em, text width = 3.3em] (b2c2d1) {\footnotesize{Anomaly Database Design~\cite{data04}}};

\node [fourthblock, below of=b3, xshift = -2.3em, yshift=-5.5em, text width = 3.3em] (b3c1d1) {\footnotesize{Time-Series Analysis for Peer Anomaly\newline~\cite{data10}}};
\node [fourthblock, below of=b3, xshift = 2.3em, yshift=-5.5em, text width = 3.3em] (b3c2d1) {\footnotesize{Sequence Similarity for Behaviour Analysis~\cite{data02}}};

\node [fourthblock, below of=b4, xshift = -2.3em, yshift=-5.5em, text width = 3.3em] (b4c1d1) {\footnotesize{Identifying DAO attack via Autoencoders~\cite{data12}}};
\node [fourthblock, below of=b4, xshift = 2.3em, yshift=-5.5em, text width = 3.3em] (b4c1d2) {\footnotesize{Identifying DAO attack via Contract Bugs~\cite{data13}}};

\node [fourthblock, below of=b5, xshift = -2.3em, yshift=-5.5em, text width = 3.3em] (b5c1d1) {\footnotesize{Tx Edge Detection via Dominant Set Analysis~\cite{data03}}};
\node [fourthblock, below of=b5, xshift = 2.3em, yshift=-5.5em, text width = 3.3em] (b5c2d1) {\footnotesize{Fraud Pre-Prediction via Time Patterns\newline~\cite{data07}}};


\node[fifthblock, below of=b1c1d1, yshift=-6em, xshift= 14em](z1){Framework Design};
\node[fifthblock, below of=b1c1d1, yshift=-6em, xshift = 21em ](z2){Clustering};
\node[fifthblock, below of=b1c1d1, yshift=-6em, xshift = 0em](z3){Graph Analysis};

\node[fifthblock, below of=b1c1d1, yshift=-6em, xshift = 7em](z7){Semi-supervised ML};

\node[fifthblock, below of=b1c1d1, yshift=-6em, xshift = 42em](z4){Multiple Machine Learning};
\node[fifthblock, below of=b1c1d1, yshift=-6em, xshift = 28em](z5){Deep Learning};
\node[fifthblock, below of=b1c1d1, yshift=-6em, xshift = 35em](z6){Ensemble Learning};


\path [line] (a1)-- ($(a1.south)+(0,-0.25)$) -|(b1);
\path [line] (a1)-- ($(a1.south)+(0,-0.25)$) -|(b2);
\path [line] (a1)-- ($(a1.south)+(0,-0.25)$) -|(b3);
\path [line] (a1)-- ($(a1.south)+(0,-0.25)$) -|(b4);
\path [line] (a1)-- ($(a1.south)+(0,-0.25)$) -|(b5);


\path [line] (b1)-- ($(b1.south)+(0,-0.25)$) -|(b1c1d1);
\path [line] (b1)-- ($(b1.south)+(0,-0.25)$) -|(b1c1d2);
\path [line] (b1)-- ($(b1.south)+(0,-0.25)$) -|(b1c2d1);

\path [line] (b2)-- ($(b2.south)+(0,-0.25)$) -|(b2c1d1);
\path [line] (b2)-- ($(b2.south)+(0,-0.25)$) -|(b2c2d1);

\path [line] (b3)-- ($(b3.south)+(0,-0.25)$) -|(b3c1d1);
\path [line] (b3)-- ($(b3.south)+(0,-0.25)$) -|(b3c2d1);

\path [line] (b4)-- ($(b4.south)+(0,-0.25)$) -|(b4c1d1);
\path [line] (b4)-- ($(b4.south)+(0,-0.25)$) -|(b4c1d2);

\path [line] (b5)-- ($(b5.south)+(0,-0.25)$) -|(b5c1d1);
\path [line] (b5)-- ($(b5.south)+(0,-0.25)$) -|(b5c2d1);



\path [line, color=cyan] (b1c1d1.south)--(z3.north);

\path [line, color=cyan] (b1c1d1.south)--(z2.north);

\path [line, color=cyan] (b1c1d2.south)--(z3.north);
\path [line, color=violet] (b1c1d2.south)--(z7.north);
\path [line, color=red] (b1c2d1.south)--(z2.north);
\path [line, color=red] (b1c2d1.south)--(z4.north);

\path [line, color=violet] (b2c1d1.south)--(z1.north);
\path [line, color=red] (b2c2d1.south)--(z2.north);

\path [line, color=orange] (b3c1d1.south)--(z5.north);

\path [line, color=red] (b3c2d1.south)--(z2.north);

\path [line, color=purple] (b4c1d1.south)--(z6.north);
\path [line, color=green] (b4c1d2.south)--(z4.north);

\path [line, color=red] (b5c1d1.south)--(z2.north);
\path [line, color=red] (b5c2d1.south)--(z2.north);

\end{tikzpicture}
	\small \caption{Classification of Blockchain Anomalies from Perspective of Data Layer}
     \label{fig:datalayer}
\end{figure*}
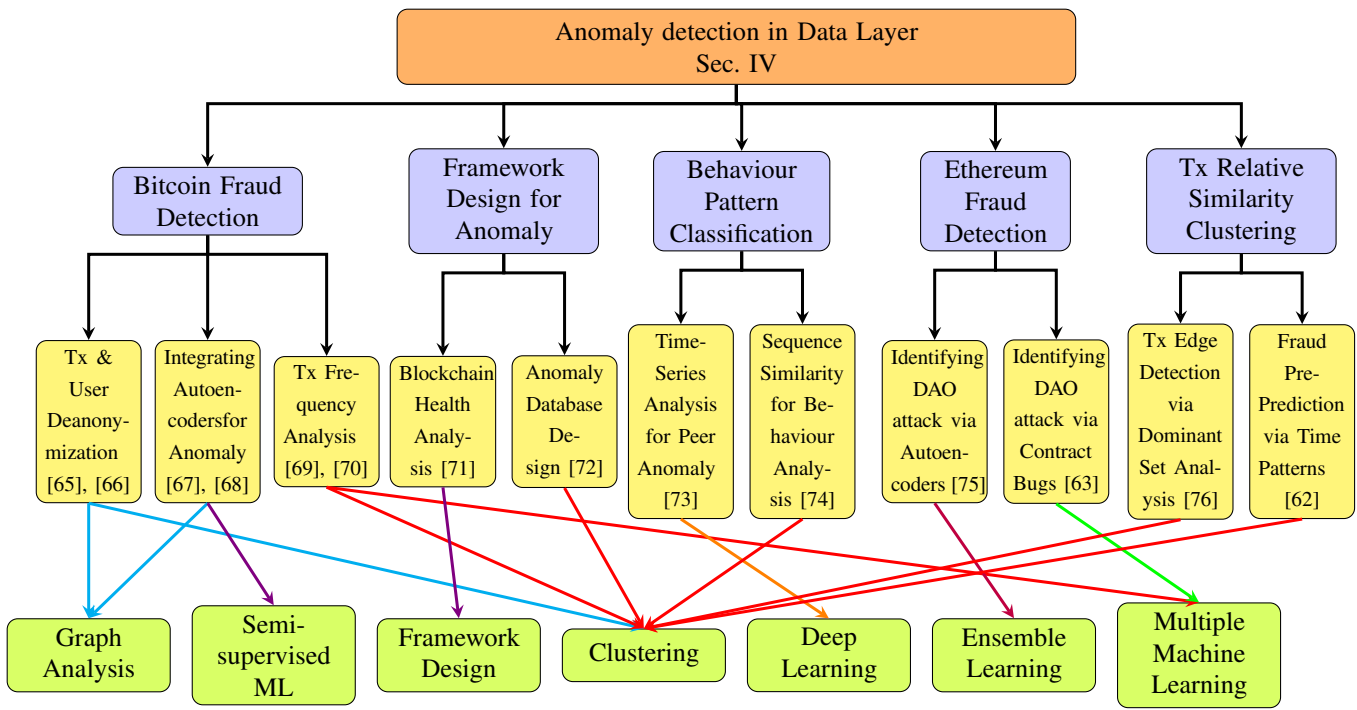

In this section, we classify the works that have highlighted these anomalies and worked over their efficient prediction from perspective of data layer (cf. Section~\ref{BlockLayers} for details)~(cf. Fig.~\ref{fig:datalayer} and Table.~\ref{tab:datalayer}).


\begin{table*}[ht]
\begin{center}
 \centering
 \scriptsize
  \captionsetup{labelsep=space}
 \captionsetup{justification=centering}
 \caption{\textsc{\\Anomaly Detection in Data Layer.}}
  \label{tab:datalayer}
  \begin{tabular}{|P{1.25cm}|P{0.55cm}|P{2.5cm}|P{2.2cm}|P{2.0cm}|P{1cm}|P{1cm}|P{1.8cm}|P{1.1cm}|P{0.6cm}|}
  	\hline
\rule{0pt}{2ex}
\bfseries Domain & \bfseries Ref No. & \bfseries Contribution & \bfseries Detected Anomaly & \bfseries  Anomaly Factors  & \bfseries \centering Blockhain Type &\bfseries Platform \newline Language & \bfseries \comst{Applications} & \bfseries Dataset & \bfseries Compl-\newline exity \\
\hline

\multirow{5}{*}{\parbox{2cm}}
\rule{0pt}{2ex}
 & \cite{data01}  & Developed a system for visual analysis of Bitcoin Flow & \tabitem Malicious Tx \newline \tabitem Pure \& impure circulated money  & \tabitem Budget \tabitem Purity \tabitem Transfer Analysis & Public & Python & \tabitem Cryptocurrency & Bitcoin Database & $-$\\
\cline{2-10}

\rule{0pt}{2ex}
& \cite{data09} & Address similarity mapping via Euclidean space  & \tabitem Anomalous Bitcoin Users & \tabitem K-similar addresses & Public & N/S & \tabitem Cryptocurrency  & Bitcoin Database & $-$ \\
\cline{2-10}

\rule{0pt}{2ex}
\centering \textbf{Bitcoin Fraud Detection} & \cite{data08} & Feature based identification of Bitcoin mixing-demixing & \tabitem Graph intermediate point for mixing services & \tabitem Tx graph reconstruction \newline \tabitem Outlier \& clustering & Public & N/S & \tabitem Cryptocurrency  & Bitcoin Database & $O(n^2)$ \\
\cline{2-10}

\rule{0pt}{2ex}
 & \cite{newref08} & \mubcom{Profiling based anomalous pattern detection from multi-dimensional data} & \tabitem Anomalous Tx Patterns & \tabitem Network traffic statistics \newline \tabitem User and Tx profiling & Public & Python & \tabitem Cryptocurrency  & Bitcoin Database & $O(n^2)$ \\
\cline{2-10}

\rule{0pt}{2ex}
& \cite{data05} & Tx pattern analysis for anomalous activity in HYIP & \tabitem Uneven payback rate \newline \tabitem Uneven Tx frequency & \tabitem Address clustering \newline \tabitem Feature gain & Public & R & \tabitem Cryptocurrency  & Bitcoin Database & $-$ \\
\cline{2-10}

\rule{0pt}{2ex}
& \cite{data11} & High order Tx moment based anomaly detection & \tabitem Malicious Tx records & \tabitem Tx moments \newline \tabitem Tx history summary & Public  & Python & \tabitem Cryptocurrency \newline \tabitem Crowdsensisng \newline \tabitem Stocks & Stock Trading Data & $-$ \\
\cline{2-10}

\hline

\rule{0pt}{2ex}
\multirow{2}{*}{\parbox{2cm}{}}
\rule{0pt}{2ex}
\centering \textbf{Framework Design} & \cite{data06} & Analysing Tx \& block interval to measure healthiness \& Anomaly of IoT blockchain &  \tabitem False IoT data storage  & \tabitem Block No \& Tx interval &Public & N/S & \tabitem IoT & Real-Time IoT Data  & $-$\\
\cline{2-10}

\rule{0pt}{2ex}
& \cite{data04} & A scalable data analysis tool design for Blockchain via MySQL and MongoDB Databases & \tabitem Uneven Tx rate & \tabitem Tx fee \newline \tabitem Address tags & Public & Multiple & \tabitem Cryptocurrency  & Bitcoin, Ethereum & $-$\\
\cline{2-10}
\hline

\multirow{2}{*}{\parbox{2cm}{}}
\rule{0pt}{2ex}
\centering \textbf{Behaviour Pattern Classification} & \cite{data10} & Classified  peers of blockchain w.r.t their behaviour via Deep Learning & \tabitem Non-similar peers  & \tabitem Batch size \newline \tabitem Class label prediction & Public  & N/S & \tabitem Cryptocurrency  & Bitcoin Database & $-$ \\
\cline{2-10}

\rule{0pt}{2ex}
& \cite{data02} & Clustering based behaviour analysis for blockchain nodes & \tabitem Anomalous behaviour sequences & \tabitem Sequence similarity & Public/\newline Private  & N/S & \tabitem Cryptocurrency \newline \tabitem Stocks \newline \tabitem IoT & Stock Trading Data & $-$ \\
\cline{2-10}
\hline

\multirow{2}{*}{\parbox{2cm}{}}
\rule{0pt}{2ex}
\centering \textbf{Ethereum Fraud Detection} & \cite{data12} & Strengthening encoder-decoder model against DAO attacks & \tabitem Decentralized autonomous organization  & \tabitem Block size \newline \tabitem Average gas usage & Public & N/S & \tabitem Cryptocurrency \newline \tabitem Non-trusted organizations & Real \& Synthetic Eth Data & $-$ \\
\cline{2-10}

\rule{0pt}{2ex}
& \cite{data13} & Malicious Tx behaviour detection via supervised learning & \tabitem Malicious Ethereum nodes  & \tabitem Tx gas analysis \newline \tabitem Tx timestamp  & Public  & Python & \tabitem Cryptocurrency  & Ethereum Data & $-$ \\
\cline{2-10}
\hline

\multirow{2}{*}{\parbox{2cm}{}}
\rule{0pt}{2ex}
\centering \textbf{Tx Relative Similarity Clustering} & \cite{data03} & Identified common behavioural nodes via dominant set analysis & \tabitem Uneven Tx behaviour  & \tabitem Tx edges \newline \tabitem cluster edges \newline \tabitem similar dominant set & Public & MATLAB & \tabitem Cryptocurrency \newline \tabitem Crowdsensing & N/S & $-$ \\
\cline{2-10}

\rule{0pt}{2ex}
& \cite{data07} & Identifying hidden time patterns from Blockchain Tx & \tabitem Anomalous Tx \newline \tabitem Anomalous behaviour nodes  & \tabitem Tx logs \newline \tabitem Levenshtein distance & Public & N/S & \tabitem Cryptocurrency  & N/S & $-$ \\
\cline{2-10}
\hline

 \end{tabular}
  \end{center}
\end{table*}


\subsection{Bitcoin Fraud Detection}
\rev{Discussing data, one cannot undermine the data available on the Bitcoin platform, which is the highest ranked cryptocurrency till now according to its market value.} The price of Bitcoin has increased dramatically in the past few years, according to a report by Investopedia, the net worth of circulating Bitcoin is more than \$600 billion in total as of May 2021~\cite{datalayer02}. On one hand, Bitcoin provides a trusted and secure medium of financial transactions, \rev{but on the other hand due to its pseudonymity it also provides a safe passage for exchange or purchase of illegal assets, services, and goods.} Therefore, it is important to highlight and take appropriate action against such activities in a timely manner.\\ 
A work by Battista~\textit{et al.} in~\cite{data01} proposed a system named as ‘BitConeView’ for analysis of Bitcoin transactions in a visual manner. The aim of the work is to carry out deanonymization of transaction flow by developing a visually analysable system. \textit{BitCoveView} allows analysing users to track the sources, flow, and patterns of Bitcoin transactions in a detailed manner. \rev{One of the critical use cases of these types of systems is to detect fraudulent and money laundering transactions,} which authors investigated deeply and claimed that their proposed model can help in detecting money laundering patterns in an efficient manner. \rev{Since the proposed framework is only suitable for Bitcoin, however, it will not be wrong to say that it can be modified and can be made suitable to other crypto-currencies following similar functioning styles. Another similar \final{article that focuses} on deanonymization and identification of anomalous users in the bitcoin network has been carried out by Shao~\textit{et al.} in~\cite{data09}. The article proposed a novel mapping based system which learns on the basis of address similarity from the perspective of a compact Euclidean space. The work identifies k-similar addresses and then uses the proposed model to identify presence of an anomalous participant. Authors carried out their experimentation on the Bitcoin database to identify anomalous nodes in the network.}\\
\rev{From the perspective of blockchain feature analysis, a unique work has been presented by Nan and Tao in~\cite{data08}. The authors focused mainly on detecting mixing and demixing services for Bitcoin cryptocurrency via transaction graph reconstruction.} By using the features of graph embedding, authors proposed a mixing identification model in which one can figure out services on the basis of their specified features. \rev{One more considerable work from the perspective of unseen patterns detection from blockchain network traffic has been presented by Kim~\textit{et al.} in~\cite{newref08}.} The work is slightly different from traditional works, as it highlights the use of network traffic rather than stored ledger data. In the proposed model, authors developed an engine which collects multi-dimensional data streams in an organized and periodic manner. The collected data is then fed to a semi-supervised learning model, which detects novel patterns from the blockchain data. The work further highlights that they introduced a profiling-based engine for efficient anomaly detection, which is implemented over autoencoder. The presented model is further tested and compared with other similar models, such as DNN, LR, GB, OC-SVM, and RF, and \rev{it can be visualized that it outperformed other models in terms of training time and detection. As model works over semi-supervised learning paradigm, it will limit identification of certain novel anomalies in the network which has not be identified before.}\\
Another critical work to identify high yielding programs for Bitcoin investments has been presented by Toyoda~\textit{et al.} in~\cite{data03}. Authors devised certain Bitcoin features and then ranked the transactions on the basis of these features to identify specific actions. To elaborate it further, authors distributed the Bitcoin payback into different distribution classes and \final{identified whether} the payback amount is from a high yielding investment program or it belongs to some other category, such as \rev{donation, exchange, mining pool, faucet, etc. In this way, the researchers were able to successfully identify and classify the malicious behaviors in order to protect Bitcoin users.}  Similar to identification of high yielding investment programs via analysing Bitcoin transactions, another article targeting analysis of transaction history for address classification has been carried out by Lin~\textit{et al.} in~\cite{data11}. \rev{The article works over proposing novel features to develop abnormality detection classification models for Bitcoin.} Authors further used these features to carry out training of supervise machine learning models, which are then used to carry out prediction and evaluation of anomalous Bitcoin addresses. \rev{From the viewpoint of Bitcoin fraud detection, the majority of works used graph analysis or traditional machine learning models. Since Bitcoin is a huge database and an abundant amount of data is available for training and testing purposes, therefore, we believe that advanced deep learning approaches should be used to enhance the accuracy.}

\subsection{Generic Framework Design for Blockchain Anomaly}
\rev{Since, blockchain is a well-applied field and now it has applications in almost every aspect of life ranging from healthcare to smart grid.} Therefore, apart from developing models just for Bitcoin transactions, \rev{it is also important to design models to check health and anomaly presence of generic blockchain models as well. One such work from the perspective of designing} of a visualization tool for anomaly detection in blockchain based IoT systems has been presented by Song~\textit{et al.} in~\cite{data06}. The proposed framework has two major aims, determination of health and detection of anomalies in blockchain networks. From the viewpoint of health classification of blockchain, authors analysed height, number of transactions, and generation interval of each block in the network. \rev{Similarly, to identify malicious activity on blockchain, authors used the data \final{of specific} number of events alongside IoT data statistics, which helped visualize and identify a prospective anomalous node. Authors used various blockchain based factors, such as block number, transaction interval, etc. to strengthen their anomaly data analysis. A more generalized work towards development of anomaly detection tools for generic blockchain networks has been presented by authors in~\cite{data04}. The authors first discuss the generic model of blockchain and then highlight and expose certain anomalies,} such as anomalous metadata, transaction fees, and address tags. Afterwards, authors implemented and validated their framework on Ethereum and Bitcoin data, which are two major blockchain models nowadays. The work developed APIs and used MongoDB and MySQL databases to evaluate their claims. \rev{After careful analysis of proposed frameworks, it will not be wrong to say that there is still a dire need to develop real-time anomaly detection frameworks for timely detection of anomalous patterns in the network.} 

\subsection{Behaviour Pattern Classification}
\rev{A blockchain network usually comprises of a large number of peer to peer nodes, which are linked with each other in a distributed decentralized manner.} This number is even more abundant in public blockchains, where anyone from any part of the world can join with no or very minimal verification. Thus, ensuring legitimacy in this large group of nodes is fairly challenging, as it is hard to classify if some nodes start misbehaving. Thus, one needs to check and classify behaviour of each participating node in order to get deeper insights and prediction about anomalous blockchain nodes.\\
A pioneering work in the field of peer behaviour classification has been carried out by Tang~\textit{et al.} in~\cite{data10}. \rev{The authors first developed a strong motivation for why traditional anomaly detection approaches such as decision tree \& SVM are not reliable and do not produce satisfactory outcomes. Afterwards, the authors proposed their own time series analysis based deep learning behaviour classification approach and named it as~\textit{PeerClassifier}. The two key factors which authors took into account while analysing behavior are batch size and class labels for the collected data.} From the experimental analysis and evaluation, authors demonstrated that their proposed approach shows significant improvements in accuracy as compared to other traditional learning approaches. Another similar work from the perspective of clustering for anomalous behaviour classification has been carried out by Huang~\textit{et al.} in~\cite{data02}. \rev{Authors first worked over evaluating the similarity list among blockchain peers and then carried out peer identification according to distance among them. Authors further evaluated their work from the perspective of precision and compared it with classical approaches to show the improvements.} An evaluation of both of the above works \final{shows that} both clustering and deep learning models provide significant improvement in anomaly detection as compared to traditional approaches. \rev{However, this field of behaviour pattern classification is not much discussed in research, and it has a lot of research potential for the future works.}
  
\subsection{Ethereum Fraud Detection}
Ethereum is the second largest blockchain platform after Bitcoin, but Ethereum is different from Bitcoin as in Ethereum, users can carry out deployment of decentralized applications alongside doing cryptocurrency transactions~\cite{datalayer04}. Furthermore, in order to deploy these decentralized applications, Ethereum provides its users the facility of decentralized smart contracts, which can be termed as pieces of code executed in a decentralized manner~\cite{datalayer05}.\\ Due to these large numbers of features and benefits, Ethereum is a big attraction for anomalous peers, as they continuously try to take unfair advantage of these features. 
\rev{The first work discussing the detection of vulnerabilities in Ethereum blockchain network by the use of ensemble deep learning has been carried out by authors in~\cite{data12}. Authors work over strengthening the encoder-decoder model for any prospective attack.} Authors did so by applying phenomenon of learning and aggregation iteratively at multiple instances, in order to carry out computation of any prospective outlier for every observed reading. \rev{Although the work did not provide a comprehensive in-depth theoretical analysis of the model, the evaluation results show that the proposed model predicts DAO attack in an efficient manner.} The second work from the perspective of detection of malicious Ethereum accounts via supervised learning approach has been carried out by Kumar~\textit{et al.} in~\cite{data13}. \rev{The major focus of the work is to understand behaviour of transactions among Ethereum accounts by keeping an eye over gas usage and transaction time-stamps.} Authors further classified the Ethereum network into two subtypes named as smart contract accounts and externally owned accounts. After this classification, authors \final{studied anomalies} in both of these types using supervised machine learning approaches such as random forest, decision free, K-NN, etc. From the evaluated outcomes, it can be seen that the proposed strategy efficiently helps in detection of anomalies in the given conditions. \rev{Nevertheless, Ethereum has a huge database, but the anomaly models over it can be enhanced further by enhancing the training and testing accuracy.}

\subsection{Tx Relative Similarity Clustering}
Apart from traditional anomalies, it is also important to study a chain of events in blockchain transactions that has led to a harmful catastrophe. \rev{One of the work, which focuses on carrying out in-depth transactional analysis by the usage of dominant set analysis, has been performed by Awan and Cortesi in~\cite{data03}.} In the work, authors emphasized that learning the behaviour of transaction for each node is critical in identifying and predicting any current and prospective anomaly. Therefore, in order to carry out efficient behaviour learning, authors proposed a dominant set approach which categorizes each transaction in the blockchain network according to the most relevant set. \rev{In order to do so, authors analysed transaction edges, and cluster edges formed as a result of analysis.} Another innovative work that emphasized over discovering  hidden time patterns via clustering in a decentralized blockchain network has been presented by authors in~\cite{data07}. The major focus of the article is to carry out future predictions via analysing current timestamp patterns in the transactions.\rev{Authors first clustered all transactions on the basis of allocated patterns, such as transaction logs and Levenshtein distance, and afterwards worked over detection anomalous behaviours by observing various parameters, such as distance, etc. Till now, both the works focusing over relative similarity of transactions use clustering, but it will not be wrong to say that this field can be explored with the help of other machine/deep learning models.}

\subsection{Summary and Insights}
\mubcom{The role of data layer in successful functioning of blockchain cannot be undermined as it acts as a backbone of blockchain from the perspective of handling and securing data records for blockchain networks. However, on the other hand, the anomalies in the data layer cannot be ignored as well, because they can cause catastrophic consequences otherwise. The anomalies over the data layer can be divided into five subtypes, \rev{in which a major proportion is occupied by the anomalies from the perspective of Bitcoin and Ethereum fraud detection. The remaining anomalies in the data layer are oriented towards finding patterns \final{of anomalous} behaviours and transactions carried out by blockchain nodes, which have been recorded over blockchain ledger.} It is important to identify such anomalies and their corresponding user accounts, so that one can restrict such accounts from carrying out such acts in the future transactions.}


\begin{figure*}[]
     \centering

\begin{tikzpicture}


\node [firstblock,  text centered, minimum width = 25em,  text width=25em] (a1) {Anomaly detection in Network Layer \\ Sec.~\ref{NetworkLayerSection} };


\node[secondblock, below of=a1, yshift=-2em, xshift=-18em, text width = 8em](b1){Malicious Accounts on Network};
\node[secondblock, below of=a1, yshift=-2em, xshift = -4.8em, text width =6em ](b2){Malicious Forks in Network};
\node[secondblock, below of=a1, yshift=-2em, xshift = 6.35em, text width = 7em](b3){Anomalous Network Transactions};
\node[secondblock, below of=a1, yshift=-2em, xshift = 17.5em, text width = 6em](b4){Network Entropy Detection};



\node [fourthblock, below of=b1, xshift = -6.8em, yshift=-5em, text width = 3.3em] (b1c1d1) {\footnotesize{Malicious User Identification from Node Graph~\cite{network03}}};
\node [fourthblock, below of=b1, xshift = -2.3em, yshift=-5em, text width = 3.3em] (b1c2d1) {\footnotesize{Makov Logic for Outliers Identification~\cite{network04}}};
\node [fourthblock, below of=b1, xshift = 2.0em, yshift=-5em, text width = 3.3em] (b1c3d1) {\footnotesize{In-Degree Scam Detection~\cite{network05}}};
\node [fourthblock, below of=b1, xshift = 6.4em, yshift=-5em, text width = 3.3em] (b1c4d1) {\footnotesize{Gini Impurity based Address Relations~\cite{network12}}};

\node [fourthblock, below of=b2, xshift = -2.25em, yshift=-5em, text width = 3.3em] (b2c1d1) {\footnotesize{Malicious Request Identification\newline~\cite{network06, network11}}};
\node [fourthblock, below of=b2, xshift = 2.25em, yshift=-5em, text width = 3.3em] (b2c2d1) {\footnotesize{Divergent Paths to Detect MI Linkage\newline~\cite{network09}}};

\node [fourthblock, below of=b3, xshift = -4.4em, yshift=-5em, text width = 3.3em] (b3c1d1) {\footnotesize{Malicious Tx Patterns via Gas Usage\newline~\cite{network02}}};
\node [fourthblock, below of=b3, xshift = -0em, yshift=-5em, text width = 3.3em] (b3c2d1) {\footnotesize{Majority Attack Resisting for Stake Holders~\cite{network08}}};
\node [fourthblock, below of=b3, xshift = 4.4em, yshift=-5em, text width = 3.3em] (b3c3d1) {\footnotesize{Time-Series Analysis for Malicious Tx~\cite{network10}}};

\node [fourthblock, below of=b4, xshift = -2.2em, yshift=-5em, text width = 3.3em] (b4c1d1) {\footnotesize{Local Outlier Factor for Tx Edges~\cite{network01}}};
\node [fourthblock, below of=b4, xshift = 2.2em, yshift=-5em, text width = 3.3em] (b4c2d1) {\footnotesize{High Speed Detection of Key Theft~\cite{network07}}};


\node[fifthblock, below of=b1c1d1, yshift=-7em, xshift = 20em, text width=4.5em ](z1){Clustering};
\node[fifthblock, below of=b1c1d1, yshift=-7em, xshift = 8em, text width=4.5em](z2){Graph Analysis};
\node[fifthblock, below of=b1c1d1, yshift=-7em, xshift = 02em, text width=4.5em](z3){SVM};
\node[fifthblock, below of=b1c1d1, yshift=-7em, xshift = 26em, text width=4.5em](z4){Kibara Tool};
\node[fifthblock, below of=b1c1d1, yshift=-7em, xshift = 14em, text width=4.5em](z5){AdVise Framework};
\node[fifthblock, below of=b1c1d1, yshift=-7em, xshift = 32em, text width=4.5em](z6){General Machine Learning};
\node[fifthblock, below of=b1c1d1, yshift=-7em, xshift = 38em, text width=4.5em](z7){Algorithmic Game Theory};
\node[fifthblock, below of=b1c1d1, yshift=-7em, xshift = 44em, text width=4.5em](z8){Rolling Window};


\path [line] (a1)-- ($(a1.south)+(0,-0.25)$) -|(b1);
\path [line] (a1)-- ($(a1.south)+(0,-0.25)$) -|(b2);
\path [line] (a1)-- ($(a1.south)+(0,-0.25)$) -|(b3);
\path [line] (a1)-- ($(a1.south)+(0,-0.25)$) -|(b4);


\path [line] (b1)-- ($(b1.south)+(0,-0.25)$) -|(b1c1d1);
\path [line] (b1)-- ($(b1.south)+(0,-0.25)$) -|(b1c2d1);
\path [line] (b1)-- ($(b1.south)+(0,-0.25)$) -|(b1c3d1);
\path [line] (b1)-- ($(b1.south)+(0,-0.25)$) -|(b1c4d1);

\path [line] (b2)-- ($(b2.south)+(0,-0.25)$) -|(b2c1d1);
\path [line] (b2)-- ($(b2.south)+(0,-0.25)$) -|(b2c2d1);

\path [line] (b3)-- ($(b3.south)+(0,-0.25)$) -|(b3c1d1);
\path [line] (b3)-- ($(b3.south)+(0,-0.25)$) -|(b3c2d1);
\path [line] (b3)-- ($(b3.south)+(0,-0.25)$) -|(b3c3d1);

\path [line] (b4)-- ($(b4.south)+(0,-0.25)$) -|(b4c1d1);
\path [line] (b4)-- ($(b4.south)+(0,-0.25)$) -|(b4c2d1);



\path [line, color=red] (b1c1d1.south)--(z1.north);
\path [line, color=orange] (b1c1d1.south)--(z3.north);

\path [line, color=green] (b1c2d1.south)--(z6.north);
\path [line, color=darkgray] (b1c3d1.south)--(z2.north);
\path [line, color=red] (b1c4d1.south)--(z1.north);

\path [line, color=brown] (b2c1d1.south)--(z5.north);
\path [line, color=green] (b2c1d1.south)--(z6.north);
\path [line, color=brown] (b2c2d1.south)--(z5.north);

\path [line, color=purple] (b3c1d1.south)--(z4.north);
\path [line, color=green] (b3c1d1.south)--(z6.north);

\path [line, color=green] (b3c2d1.south)--(z6.north);
\path [line, color=violet] (b3c2d1.south)--(z7.north);
\path [line, color=blue] (b3c3d1.south)--(z8.north);

\path [line, color=red] (b4c1d1.south)--(z1.north);
\path [line, color=red] (b4c2d1.south)--(z1.north);

\end{tikzpicture}
	\small \caption{Classification of Blockchain Anomalies from Perspective of Network Layer}
     \label{fig:networklayer}
\end{figure*}
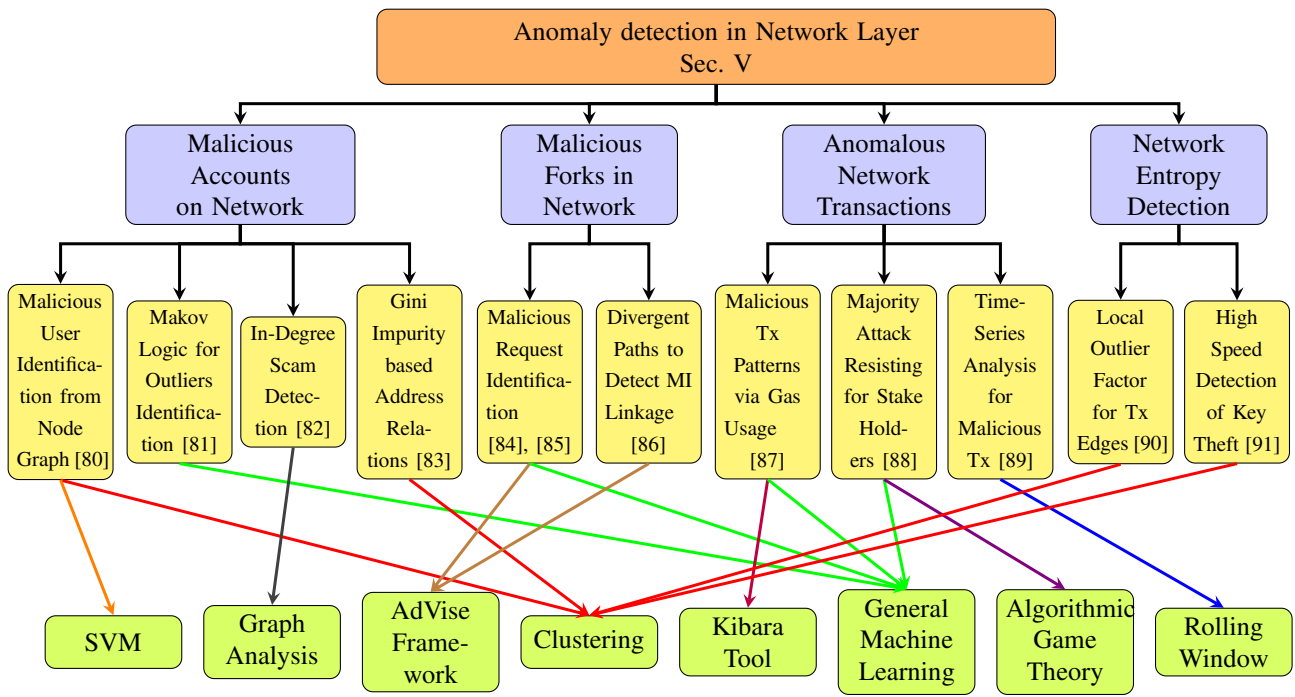



\begin{table*}[ht]
\begin{center}
 \centering
 \scriptsize
  \captionsetup{labelsep=space}
 \captionsetup{justification=centering}
 \caption{\textsc{\\Anomaly Detection in Network Layer.}}
  \label{tab:networklayer}
  \begin{tabular}{|P{1.25cm}|P{0.55cm}|P{2.5cm}|P{2.2cm}|P{2.0cm}|P{1cm}|P{1cm}|P{1.8cm}|P{1.1cm}|P{0.6cm}|}
  	\hline
\rule{0pt}{2ex}
\bfseries Domain & \bfseries Ref No. & \bfseries Contribution & \bfseries Detected Anomaly & \bfseries  Anomaly Factors  & \bfseries \centering Blockhain Type &\bfseries Platform \newline Language & \bfseries \comst{Applications} & \bfseries Dataset & \bfseries Compl-\newline exity \\
\hline

\multirow{2}{*}{\parbox{2cm}}
\rule{0pt}{2ex}
 & ~\cite{network03}  & Evaluated three unsupervised learning models for Bitcoin anomaly & \tabitem Malicious users \newline \tabitem Malicious Tx & \tabitem Graph overlapping  \newline \tabitem Out-degree & Public & N/S & \tabitem Cryptocurrency  & Bitcoin Database  & $-$ \\
\cline{2-10}

\rule{0pt}{2ex}
\centering \textbf{Malicious Accounts\newline on Network} & ~\cite{network04} & Random Forest algorithm based prediction for ground truth cases & \tabitem Fraudulent nodes & \tabitem Precision \& Recall of Tx users data & Public & Python  & \tabitem Cryptocurrency \newline \tabitem Markov Logic Networks & Bitcoin Database & $-$ \\
\cline{2-10}

\rule{0pt}{2ex}
& ~\cite{network05} & Exploited structural properties of graph to find unusual patterns & \tabitem Artificial Tx patterns & \tabitem Common \& Uncommon Output Amount \newline \tabitem Tx Frequency & Public & N/S & \tabitem Cryptocurrency & Bitcoin Database & $-$ \\
\cline{2-10}

\rule{0pt}{2ex}
& \cite{network12} & Analyzed Bitcoin Tx patterns via clustering of ownership records & \tabitem Fraudulent owner clusters & \tabitem Gini Impurity Index & Public & Blockseer & \tabitem Cryptocurrency  & Bitcoin & $-$ \\
\cline{2-10}

\hline

\rule{0pt}{2ex}
\multirow{3}{*}{\parbox{2cm}{}}

\rule{0pt}{2ex}
& \cite{network06} & Used meta-data of blockchain systems to tackle eclipse attacks & \tabitem Malicious requests \newline \tabitem Malicious forks & \tabitem Pattern requests & Public & N/S & \tabitem Cryptocurrency \newline \tabitem IoT  & N/S & $-$ \\
\cline{2-10}

\rule{0pt}{2ex}
\centering \textbf{Malicious Forks in Network} & \cite{network11} & Developed a novel Blockchain anomaly detection system & \tabitem Malicious code \newline \tabitem Malicious requests & \tabitem Bandwidth overhead  \newline \tabitem Request Patterns & Public & N/S & \tabitem Cryptocurrency \newline \tabitem IoT & Bitcoin Dataset & $O(k)$ \\
\cline{2-10}

\rule{0pt}{2ex}
& \cite{network09} & Link-mining tool based anomaly detection for IoT & \tabitem Malicious forks & \tabitem Mutual Information & Public & N/S & \tabitem Cryptocurrency \newline \tabitem IoT  & N/S & $-$ \\
\cline{2-10}
\hline

\rule{0pt}{2ex}
\multirow{3}{*}{\parbox{2cm}{}}

\rule{0pt}{2ex}
& \cite{network02} & Using visualized features to detect anomalous gas spent & \tabitem Malicious Tx & \tabitem Tx throughput \newline \tabitem Gas usage & Public/ \newline Private & Node.js \newline Python & \tabitem Cryptocurrency \newline \tabitem DApps & Ethereum & $-$ \\
\cline{2-10}

\rule{0pt}{2ex}
\centering \textbf{Anomalous Network Transactions} & \cite{network08} & Stakeholder activity monitoring via software agents & \tabitem Malicious nodes \newline \tabitem Double spending & \tabitem Tx Payoff & Public & N/S & \tabitem Cryptocurrency & Bitcoin Database & $-$ \\
\cline{2-10}

\rule{0pt}{2ex}
& \cite{network10} & Personalized detection of anomaly via automated Tx signing & \tabitem Malicious Tx & \tabitem Tx time-frame \newline \tabitem Tx frequency & Public & Python & \tabitem Cryptocurrency & Ethereum & $-$ \\
\cline{2-10}
\hline

\rule{0pt}{2ex}
\multirow{3}{*}{\parbox{2cm}{}}

\rule{0pt}{2ex}
\centering \textbf{Malicious Forks in Network} & \cite{network01} & Local outlier factor based clustering for anomaly detection & \tabitem Suspicious Tx \newline \tabitem Suspicious users & \tabitem Tx Edges & Public & N/S & \tabitem Cryptocurrency & Bitcoin Dataset & $-$ \\
\cline{2-10}

\rule{0pt}{2ex}
& \cite{network07} & High-Speed anomaly detection for blockchain using In-GPU cache & \tabitem Suspicious Tx & \tabitem Abnormal execution time \newline \tabitem Avg withdrawal \& deposit & Public & CUDA & \tabitem Cryptocurrency & Bitcoin Dataset & $-$ \\
\cline{2-10}
\hline

 \end{tabular}
  \end{center}
\end{table*}


\section{Anomaly Detection in Network Layer of Blockchain} \label{NetworkLayerSection}

In this section, we provide an in-depth discussion about the functionalities, limitation, and comparative analysis of anomaly detection models from perspective of the network layer of blockchain (cf. Section~\ref{BlockLayers} for details)~(cf. Fig.~\ref{fig:networklayer} and Table.~\ref{tab:networklayer}).   
 
\subsection{Malicious Accounts on Network}
In a decentralized blockchain network, \rev{one of the prime focuses for anomalous peers on the network is to mask their identity so that they become untraceable.} Anomalous peers try to hide their identities by taking unfair advantages of loopholes in the network. \rev{This category of masking the identity while using benefits of the network comes under adversarial activities over the blockchain network layer, therefore, in this section, we first discuss the works which identify the identities of these malicious accounts.} One of the critical work in this regard has been carried out by Phan and Lee in~\cite{network03}. The work focused on analysing behaviours of users on the Bitcoin network by analysing network graphs via different unsupervised learning mechanisms. In order to carry out unsupervised learning on users and transactions graphs, authors used three renowned mechanisms named as support vector machine (SVM), k-Means clustering, and Mahalanobis distance. \rev{Since this work uses unsupervised learning, the effectiveness and accuracy of the proposed model is not pretty high. \rev{However, from the evaluation outcome, authors were able to identify a few cases of fraud and theft accordingly.} Another detailed work which focuses on detection of malicious behaviour on the network of Bitcoin cryptocurrency has been presented in the form of a thesis by Frank Jobse in~\cite{network04}. The work first discussed suspicious patterns and fraud detection in the Bitcoin network, and afterwards, it provides detailed discussion about dataset, data analysis, and sampling techniques used in the evaluation. \rev{After that the author presented a discussion about Markov logic network's usage in the anomaly detection methodology, and finally the work evaluated the proposed model and compared the work with baseline methods to show the effectiveness from the viewpoint of precision.}} \\
\rev{One more work from the perspective of detection of artificial and strange behaving nodes in the network via user graphs in Bitcoin has been performed by Maesa~\textit{et al.} in~\cite{network05}.} The article majorly emphasizes over the outliers in the category of indegree distribution of frequency and remarkably high diameters. Article further discussed the formation of various chain transactions by providing in-depth discussion about various transaction types, such as ~\textit{BPS, GPS, PS}- transactions. \rev{After that, the article evaluated the economic meaning of these transactions and related the anomalous output behavior to these transaction types for successful identification of malicious nodes. The final work in this domain of malicious account detection on the network layer of blockchain has been carried out by Chang and Svetinovic in~\cite{network12}.} Authors worked over analysing different transaction patterns with a goal to cluster the addresses with similar ownership information. In order to do so, authors developed a clustering approach and clustered all transactions on the network into five different patterns such as peeling transactions, relay transactions, etc. Another novel that authors did is that they used Gini impurity measure to evaluate the outcome of the proposed clustering model. \rev{Authors further compared the distributions with normal distribution and after applying the proposed model to carry out comparative analysis for the work to show a comparative analysis. Since identifying malicious accounts is a critical challenge, thus, different types of detection models can be seen in this category, such as SVM, clustering, graph analysis, etc. However, this domain also needs further advancement to integrate recently developed models to improve detection accuracy. This can be done via multiple ways but we believe that usage of modern deep learning and reinforcement learning based models is one of them.}

\subsection{Malicious Forks in Network}
The simplest definition of fork can be termed as disagreement on choosing the best way forward for the blockchain network, this disagreement usually occurs between multiple miners, which control the computational power of blockchain network~\cite{networklayer03}. As a fork results in splitting of blockchain into two separate chains, therefore, there is a strong possibility that a fork can be carried out for both advantageous and malicious purposes~\cite{networklayer04}.\\ 
A number of works have been carried out in efficient identification and prediction of these forks, \final{one such} work in this regard has been presented by Pontecorvi~\textit{et al.} in~\cite{network06}. \rev{Authors developed a postulate that malicious requests in the network leads to the formation of malicious forks and it also paves a path for prospective attacks in the network. Therefore, in order to efficiently eradicate these catastrophic outcomes, it is important to detect and overcome these malicious requests. In order to do so, authors developed a malicious activity detection tool and named it as AdvISE. The \final{proposed tool} collects and analyses data of requests in the  blockchain networks and highlights potential adversarial requests for timely action.} The extension and further implementation and investigation of this idea has been carried out by authors in~\cite{network11}, in which authors developed a thorough anomaly detection and fork leveraging tool for blockchain named it a `BAD’. \rev{Authors developed a complete framework to tackle anomalies at a massive level, and also made the proposed system resilient from eclipse attack.} Afterwards, authors implemented a thorough test bed, in which they implemented the complete network by using two types of nodes named as full nodes and client nodes. \rev{From the evaluation viewpoint, \rev{the authors ensured that the proposed model detects anomalies and occurrences of eclipse attack in the network from analysing transaction patterns.} Another interesting work focusing over the use of link-mining tools on the basis of blockchain network anomaly detection has been presented by Agure~\textit{et al.} in~\cite{network09}.} The aim of the work is to collect blockchain meta-data in the form of network forks, which are further used to figure out prospective anomalous paths in the network. \rev{A critical parameter, which is used to aid the experimentation is mutual information (MI), which is used as a measure to ensure efficiency of the proposed model from the perspective of detection and occurrence of anomalies and forks. After analysing the current literature, it can be visualized that detection of malicious forks has not been well-studied in the literature and integration of modern machine models is still lacking in this category. Therefore, in order to further excel in this domain and to detect the formation of malicious forks in a timely manner, one needs to study their behaviors with the help of modern algorithms, such as CNN, DNN, DQN, etc. }
 
\subsection{Anomalous Tx on Network}
\rev{In the previous section (Sec.~\ref{DataLayerSection}), we discuss how anomalous transactions can be identified from the stored data records on the data layer.} However, research works have indicated that these transactions can be filtered even before recording them to the data layer of blockchain. In this section, we discuss the network layer aspects and detection of these anomalous transactions. The first work in this regard has been carried out by Bogner in~\cite{network02}. The work introduces an online solution based on machine learning for optimal visualization of anomalous transactions on the network. The goal of the experimentation is to design a user-friendly visualization tool which will be easy enough to be used by non-technical human operators to identify prospective anomalies in the network transactions. \\
Another pioneering work which aims to detect malicious transactions intended for the purpose of majority attack has been carried out by Dey in~\cite{network08}. Author first developed the motivation that in consortium blockchain, the chances of majority-attack is far greater as compared to public networks because any collusion among governing companies can result in a majority attack. Afterwards, the author discussed that the malicious transactions intended for the purpose of majority attack can be detected and prevented timely if appropriate measures are taken. \rev{In order to detect and prevent such malicious transactions, the author used game-theoretic supervised learning model, which can detect the legitimacy of a transaction and stakeholder on the basis of the past transactions.\\
Another pioneering work using machine learning for automated transaction signing to ensure efficient anomaly detection on blockchain networks has been carried out by Podgorelec~\textit{et al.} in~\cite{network10}. Authors first emphasized that digital signing of transactions takes time,} and that is the prime reason that blockchain is not being integrated in time-critical applications. Afterwards, authors work over proposing an automated and decentralized digital signing framework on the basis of machine learning, which according to the claim not only will make the blockchain efficient but will also detect anomalous transactions at the time of digital signature via time-series machine learning analysis. While evaluating the framework, authors carried out a comparison between the proposed framework and the original process and demonstrated that their proposed framework optimizes blockchain efficiency and anomaly detection. \rev{From the works, it will not be wrong to say that this indeed in a well-studied domain, as this is one of the few domains where researchers used the notion of algorithmic game theory. However, this is still the start and this domain can be extended further where one will be able to identify anomalous transactions over the network without the risk of compromising their data. E.g., via using privacy preserving machine/deep learning models for anomaly analysis.}

\subsection{Network Entropy Detection}
\rev{In fact, this section~\ref{NetworkLayerSection} as a whole section discusses anomalies and their detection in the network layer of blockchain. But works discussed prior to this subsection provide information about analysis of networks for some particular issue, such as malicious forks, transactions, accounts, etc. However, in this particular subsection, we discuss how we can make the whole network secure} from generic anomalies and what are the works that have been carried out in this domain so far. One pioneering work by Pham and Lee~\cite{network01} provides a thorough analysis about integration of \rev{anomaly detection in the network layer of Bitcoin cryptocurrency. First of all, the authors developed a methodology for data collection from the Bitcoin network, in which they classified different types of data streams in user and truncation graphs.} Afterwards, authors developed a $k$-means clustering model, which uses six features from user-node and three features from \rev{transaction nodes and clusters them accordingly. Afterwards, authors work over identification of anomalies, for which they worked from the perspective of local power degree, outlier factor, and densification laws. Another novel work from the perspective of accelerating the process of anomaly detection for blockchain networks has been carried out by authors in~\cite{network07}.} Authors first developed motivation about their work by stating the issues which can be caused due to a malicious transaction if it gets recorded on a tamper proof ledger. Thus, in order to eradicate these issues, authors mentioned that high-speed anomaly detection at the network layer is required, so that one stops malicious transactions from being recorded on the ledger. To \final{facilitate this,} authors developed a model which uses a $k$-means algorithm to detect anomalous transactions, however, to accelerate the process, authors propose a model which carry out both abnormality detection and feature extraction in GPU memory. The proposed model is then evaluated and compared with traditional models, \rev{which showed that the proposed model was 37.1 times quicker than the traditional CPU based processing model. TO demonstrate it further, authors compared it with a traditional GPU based model, which does not carry out feature extraction in GPU, the results showed that the proposed model is 16.1 times speedier than the traditional GPU based model. The works have highlighted that only clustering based models have been used to study the entropy in the network, therefore, this domain can be enhanced with the help of modern machine/deep learning models. Especially, with the help of generative networks, one can reconfigure the entropy and study its behavior.}

\subsection{Summary and Insights}
\mubcom{\rev{Network layer on blockchain is responsible for carrying out activities related} to communication and information delivery over the blockchain network. Since, this layer is establishing communication between multiple nodes and is ensuring the legitimacy of transactions and data being transferred, thus the anomalies and frauds for this layer are pretty disastrous and need strong consideration. The anomalies over the network layer of blockchain can be divided into four subtypes on the basis of their impact. The most prominent type is malicious accounts over the networks, where \rev{anomalous users try to pretend to be legitimate ones. The next type includes formation of malicious forks over the network which is done via either carrying out malicious requests over the network or via making divergent paths. The next two types consist of carrying out anomalous transactions over the network and \final{via carrying} out anomalous behaviours over the network, such as key theft, etc.} Irrespective of the type of anomaly, it is important to highlight that as these anomalies are usually being done via communication link, thus, they can be traced and stopped before causing catastrophe, if proper actions are taken.}


\begin{figure*}[]
     \centering

\begin{tikzpicture}


\node [firstblock,  text centered, minimum width = 25em,  text width=25em] (a1) {Anomaly detection in Incentive/Currency Layer \\ Sec.~\ref{IncentiveLayerSection} };


\node[secondblock, below of=a1, yshift=-2.5em, xshift=-12em, text width = 7em](b1){Bitcoin Fraud Detection};
\node[secondblock, below of=a1, yshift=-2.5em, xshift = 6em, text width = 6em ](b2){Malicious Blockchain Account};
\node[secondblock, below of=a1, yshift=-2.5em, xshift = 14em, text width = 6.5em](b3){Incentive Attack Classification};
\node[secondblock, below of=a1, yshift=-2.5em, xshift = 22em, text width = 6.5em](b4){Market Volatility Prediction};



\node [fourthblock, below of=b1, xshift = -10em, yshift=-4.5em, minimum width = 2.5mm] (b1c1d1) {\footnotesize{Atypical Tx Patterns for Coin Mixing~\cite{incentive01}}};
\node [fourthblock, below of=b1, xshift = -5em, yshift=-4.5em, minimum width = 2.5mm] (b1c2d1) {\footnotesize{Illicit Activity via F-1 Score~\cite{incentive12}}};
\node [fourthblock, below of=b1, xshift = 0em, yshift=-4.5em, minimum width = 2.5mm] (b1c2d2) {\footnotesize{Outlier Identification~\cite{incentive04}}};
\node [fourthblock, below of=b1, xshift = 5em, yshift=-4.5em, minimum width = 2.5mm] (b1c3d1) {\footnotesize{Malicious Laundering Patterns~\cite{incentive06}}};
\node [fourthblock, below of=b1, xshift = 10em, yshift=-4.5em, minimum width = 2.5mm] (b1c4d1) {\footnotesize{Attribute Indicator for Wallets~\cite{incentive02}}};

\node [fourthblock, below of=b2, xshift = -2.5em, yshift=-4.5em, minimum width = 2.5mm] (b2c1d1) {\footnotesize{Malicious Pattern via Subgraphs~\cite{incentive07}}};
\node [fourthblock, below of=b2, xshift = 2.5em, yshift=-4.5em, minimum width = 2.5mm] (b2c2d1) {\footnotesize{Malicious Feature Engineering~\cite{incentive09, incentive11}}};

\node [fourthblock, below of=b3, xshift = -0em, yshift=-4.5em, minimum width = 2.5mm] (b3c1d1) {\footnotesize{Attacks Attribute Analysis~\cite{incentive08}}};

\node [fourthblock, below of=b4, xshift = -2.5em, yshift=-4.5em, minimum width = 2.5mm] (b4c1d1) {\footnotesize{Forecasting Bitcoin Market Fluctuation~\cite{incentive03}}};
\node [fourthblock, below of=b4, xshift = 2.5em, yshift=-5em, minimum width = 2.5mm] (b4c2d1) {\footnotesize{Graphical Tx Vector Analysis~\cite{incentive10}}};


\node[fifthblock, below of=b1c1d1, yshift=-5em, xshift= 11em](z1){K-Means Clustering};
\node[fifthblock, below of=b1c1d1, yshift=-5em, xshift = 03em ](z2){Active Learning};
\node[fifthblock, below of=b1c1d1, yshift=-5em, xshift = 43em](z3){SVM};
\node[fifthblock, below of=b1c1d1, yshift=-5em, xshift = 27em](z4){Graph Analysis};
\node[fifthblock, below of=b1c1d1, yshift=-5em, xshift = 35em](z5){Decision Tree ML};
\node[fifthblock, below of=b1c1d1, yshift=-5em, xshift = 19em](z6){Random Forest};


\path [line] (a1)-- ($(a1.south)+(0,-0.25)$) -|(b1);
\path [line] (a1)-- ($(a1.south)+(0,-0.25)$) -|(b2);
\path [line] (a1)-- ($(a1.south)+(0,-0.25)$) -|(b3);
\path [line] (a1)-- ($(a1.south)+(0,-0.25)$) -|(b4);


\path [line] (b1)-- ($(b1.south)+(0,-0.25)$) -|(b1c1d1);
\path [line] (b1)-- ($(b1.south)+(0,-0.25)$) -|(b1c2d1);
\path [line] (b1)-- ($(b1.south)+(0,-0.25)$) -|(b1c2d2);
\path [line] (b1)-- ($(b1.south)+(0,-0.25)$) -|(b1c3d1);
\path [line] (b1)-- ($(b1.south)+(0,-0.25)$) -|(b1c4d1);

\path [line] (b2)-- ($(b2.south)+(0,-0.25)$) -|(b2c1d1);
\path [line] (b2)-- ($(b2.south)+(0,-0.25)$) -|(b2c2d1);

\path [line] (b3)-- ($(b3.south)+(0,-0.25)$) -|(b3c1d1);

\path [line] (b4)-- ($(b4.south)+(0,-0.25)$) -|(b4c1d1);
\path [line] (b4)-- ($(b4.south)+(0,-0.25)$) -|(b4c2d1);



\path [line, color=brown] (b1c1d1.south)--(z1.north);

\path [line, color=violet] (b1c2d1.south)--(z2.north);
\path [line, color=brown] (b1c2d2.south)--(z1.north);

\path [line, color=red] (b1c3d1.south)--(z4.north);
\path [line, color=brown] (b1c4d1.south)--(z1.north);

\path [line, color=red] (b2c1d1.south)--(z4.north);
\path [line, color=red] (b2c2d1.south)--(z4.north);
\path [line, color=orange] (b2c2d1.south)--(z5.north);
\path [line, color=blue] (b2c2d1.south)--(z6.north);

\path [line, color=magenta] (b3c1d1.south)--(z3.north);

\path [line, color=red] (b4c1d1.south)--(z4.north);
\path [line, color=red] (b4c2d1.south)--(z4.north);

\end{tikzpicture}
	\small \caption{Classification of Blockchain Anomalies from \final{the Perspective} of Incentive Layer}
     \label{fig:incentivelayer}
\end{figure*}
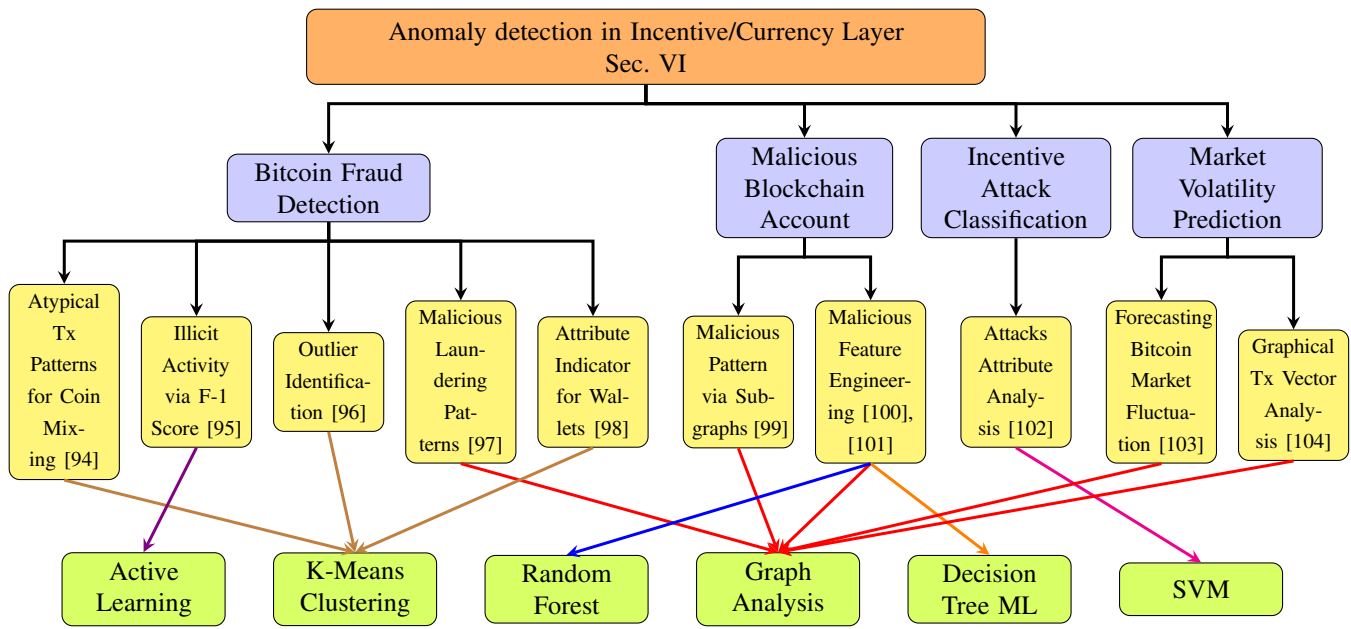


\section{Anomaly Detection in Incentive Layer of Blockchain}\label{IncentiveLayerSection}
In this section, we provide a thorough literature review from perspective of anomaly detection in incentive/currency layer of blockchain (cf. Section~\ref{BlockLayers} for details)~(cf. Fig.~\ref{fig:incentivelayer} and Table.~\ref{tab:incentivelayer}).


\begin{table*}[ht]
\begin{center}
 \centering
 \scriptsize
  \captionsetup{labelsep=space}
 \captionsetup{justification=centering}
 \caption{\textsc{\\Anomaly Detection in Incentive Layer.}}
  \label{tab:incentivelayer}
  \begin{tabular}{|P{1.25cm}|P{0.55cm}|P{2.5cm}|P{2.2cm}|P{2.0cm}|P{1cm}|P{1cm}|P{1.8cm}|P{1.1cm}|P{0.6cm}|}
  	\hline
\rule{0pt}{2ex}
\bfseries Domain & \bfseries Ref No. & \bfseries Contribution & \bfseries Detected Anomaly & \bfseries  Anomaly Factors  & \bfseries \centering Blockhain Type &\bfseries Platform \newline Language & \bfseries \comst{Applications} & \bfseries Dataset & \bfseries Compl-\newline exity \\
\hline

\multirow{5}{*}{\parbox{2cm}}
\rule{0pt}{2ex}
 & \cite{incentive01} & Clustering hub based coin mixing detection  & \tabitem Atypical Tx patterns & \tabitem Hub count \newline \tabitem Tx value variance & Public & N/S & \tabitem Cryptocurrency & Bitcoin & $-$ \\
\cline{2-10}

\rule{0pt}{2ex}
& \cite{incentive12} & Detected money laundering via supervised learning & \tabitem Illicit Tx & \tabitem F-1 Score  & Public & Python & \tabitem Cryptocurrency & Bitcoin & $-$ \\
\cline{2-10}

\rule{0pt}{2ex}
\centering \textbf{Bitcoin Fraud} & \cite{incentive04} & Detecting global \& local frauds of Bitcoins via supervised and unsupervised learning & \tabitem Malicious Tx groups & \tabitem Account inputs \newline \tabitem Account outputs & Public & N/S & \tabitem Cryptocurrency & Bitcoin & $-$ \\
\cline{2-10}

\rule{0pt}{2ex}
& \cite{incentive06} & Identifying laundering patterns from Hypergraph with high accuracy & \tabitem Malicious Tx patterns & \tabitem Exchange addresses & Public & Python & \tabitem Cryptocurrency & Bitcoin & $-$ \\
\cline{2-10}

\rule{0pt}{2ex}
& \cite{incentive02} & Analyzed Bitcoin data to detect occurred fraud & \tabitem Compromised wallets & \tabitem Tx frequency \newline \tabitem User occurrence frequency  & Public & N/S & \tabitem Cryptocurrency & Bitcoin & $-$ \\
\cline{2-10}

\hline

\rule{0pt}{2ex}
\multirow{3}{*}{\parbox{2cm}{}}

\rule{0pt}{2ex}
& \cite{incentive07} & GPU based high speed anomaly detection from user-centric subgraphs & \tabitem Abnormal Tx & \tabitem Tx Edges & Public & CUDA & \tabitem Cryptocurrency & Ethereum & $-$ \\
\cline{2-10}

\rule{0pt}{2ex}
\centering \textbf{Malicious Blockchain Accounts} & \cite{incentive09} & Random Forest based malicious node detection & \tabitem Tx with discernible purpose & \tabitem Wallet labelling & Public & Python & \tabitem Cryptocurrency & Bitcoin & $-$ \\
\cline{2-10}

\rule{0pt}{2ex}
& \cite{incentive11} & Using XGBoost classifier to detect malicious features & \tabitem Illicit accounts & \tabitem Avg Tx value \newline \tabitem Received \& sent values & Public & Python & \tabitem Cryptocurrency & Ethereum & $-$ \\
\cline{2-10}
\hline

\rule{0pt}{2ex}
\multirow{1}{*}{\parbox{2cm}{}}

\rule{0pt}{2ex}
\centering \textbf{Incentive Attack Classification} & \cite{incentive08} & Used one-class SVM and K-Means on electronic Tx data & \tabitem Malicious Tx & \tabitem Tx number, address, \& volume  & Public & Python & \tabitem Cryptocurrency & Bitcoin & $-$ \\
\cline{2-10}
\hline

\rule{0pt}{2ex}
\multirow{2}{*}{\parbox{2cm}{}}

\rule{0pt}{2ex}
\centering \textbf{Market Volatility Prediction} & \cite{incentive03} & Predicting volatility \& return in Bitcoin market via network theory & \tabitem Price variance & \tabitem Market In-Out Tx \newline \tabitem Impulse response & Public & N/S & \tabitem Cryptocurrency & Bitcoin & $-$ \\
\cline{2-10}

\rule{0pt}{2ex}
& \cite{incentive10} & Identified mining based market manipulation of Bitcoin & \tabitem Fluctuating exchange rate \newline \tabitem Abnormal Tx patterns & \tabitem Account BTC production & Public & N/S & \tabitem Cryptocurrency & Bitcoin & $-$ \\
\cline{2-10}
\hline

 \end{tabular}
  \end{center}
\end{table*}


\subsection{Bitcoin Fraud Detection}
An initial work in Bitcoin fraud identification as a part of course project has been carried out by Hirshman~\textit{et al.} in~\cite{incentive01}. The focus of the article is to define and figure out atypical transaction patterns in Bitcoin currency. \final{In the work, authors first} performed relational checks on Bitcoin data in order to figure out the roots of coin mixing for any anomalous transaction. In order to do so, authors used the K-means clustering model and developed and identified different clusters on the basis of degree variance and hub count. Finally, authors examined certain real-time splits to figure out the level of anomaly from various coin mixing services alongside identifying certain intermediate addresses involved in the malicious transaction.\\
\rev{A very interesting work from the perspective of anomaly detection in the Bitcoin network in the existence of label scarcity has been carried out by Lorenz~\textit{et al.} in~\cite{incentive12}.} The focus of the article is to basically detect money laundering patterns in cryptocurrencies by specially focusing over Bitcoin. Authors first highlighted that traditional unsupervised money laundering detection \rev{models are not good enough for the Bitcoin network, and therefore, authors designed supervised learning models to detect illicit laundering patterns in the network.} In order to evaluate the proposed model, authors worked over reporting unlawful F1-score in a unit time-step performed during the test. The reported scores are then used to identify anomalous users in order to take action against them. Another similar work that focuses over the usage of global and local outliers for the identification of Bitcoin fraud has been carried out by Monamo~\textit{et al.} in~\cite{incentive04}. \rev{The authors first highlighted the issue that lack of class labels in the Bitcoin network is one of the root causes due to which it is hard to \final{find financial} anomalies in the network. Afterwards, authors discussed fraud in the Bitcoin network from both a global and local perspective.} Then in order to identify the anomalies, authors highlighted the use of both unsupervised and supervised models for the identification of global and local outliers. For unsupervised models, authors worked over $k-means$ and $kd-trees$ clustering, in which they identified that the clustering mode of `8’ gives the optimal result. Similarly, for supervised learning models, authors used GLM logistic regression, boosted logistic regression, and random forest. Authors further emphasized the use of supervised learning models on the basis of findings and the detection accuracy of these models. \\
\rev{Till now, we discussed the use of outliers and similar other patterns, but a very different work from the perspective of using Hypergraph for malicious user identification of Bitcoin has been carried out by authors in~\cite{incentive06}. The article focuses on identification of specific exchange patterns of Bitcoin with respect to its spending and acquisition.} To study it further, authors work over building a classification model which discriminates various feature features and the major focus was to identify the root of a malicious address, which means verification of an address that whether it is \rev{owned by a specific exchange or not. The basic reason behind designing and analysing 2-motif hypergraphs to figure out hidden patterns via learning models.} To evaluate it further, authors used five learning models named as linear SVM, perceptron, random forest, logistic regression, and AdaBoost. Authors compared these models on the basis of their precision, recall, and F1 score. A final work that evaluates and proposes the anomalous aspects in Bitcoin \final{wallets has been} carried out by Zambre and Shah in~\cite{incentive02}. The developed project aims to identify the malicious users and entities who are targeting vulnerable wallets and accounts of Bitcoin users with an intention to compromise them for illicit purposes. The article gave examples of certain robberies and thefts that have been carried out over the Bitcoin network so far, and afterwards, authors used k-means clustering for malicious user identification. In order to get efficient results for this k-means clustering, authors first extracted 21 available features from available Bitcoin data and afterwards evaluated and categorised users on the basis of occurrence frequency. Authors mentioned that they were able to detect the illicit behaviour with 76.5 percent accuracy. \\
\rev{Despite the abundant number of works in the domain of Bitcoin anomaly detection in the incentive layer, it is important to mention that this domain still lacks a lot, and there is a huge need for more research in order to make the cryptocurrency more secure and trustworthy for future users. For instance, as the data records are sufficient from training, thus apart from clustering and graph analysis, researchers can work over integration of reinforcement and generative architecture based models, which can yield more accurate data.}

\subsection{Malicious Blockchain Accounts}
Since blockchain is immutable, therefore, it is also equally important to detect the malicious transactions before updating them on the ledger, and for this, we need highly efficient models that scrutinize transactions at a high pace. One such work has been carried out by Morishima in~\cite{incentive07}. The work basically revolves \final{around the} use of GPUs to speed up anomaly detection processes in blockchain networks. \rev{In the article, authors first used the concept of fixed size subgraphs, which are centred towards blockchain users in order to develop an anomaly detection model.} However, use of these types of subgraphs normally result in increase in the total time of execution for any model. Therefore, in order to overcome these execution time issues, authors work over proposing GPU oriented structural graphs which speed up the execution and detection process for a timely action. Authors further evaluated their proposed model over 300 million transactions and claimed that their proposed model provides 195 times faster execution time as compared to traditional methods. Similarly, from the perspective of detection accuracy, authors claimed that their true positive rate is substantially larger as compared to traditional anomaly detection models due to the use of GPU and developed subgraphs. \\
Apart from attacks over users’ identities, attackers and adversaries also try to play with different transactional features in order to compromise accounts so that they can steal critical assets or cryptocurrencies. Therefore, playing with different features to identify anomalies in blockchain is a critical aspect which needs more consideration. Till now, two critical works~\cite{incentive09, incentive11}, have been carried out in this domain so far, one from the perspective of malicious features for compromised wallets identification and other from the point of view of top rated feature engineering for timely user anomaly identification.  The first work in the domain has been carried out by Baek~\textit{et al.} in ~\cite{incentive09}. The article mainly focuses on investigating Binance platform, which is one of the most commonly used cryptocurrency platforms nowadays. Authors evaluated more than 38,000 wallets in order to identify transactions for malicious purposes. In order to enhance the detection, authors worked over feature engineering, in which they identified and used the \rev{most suitable features for unsupervised learning models, such as random forest.} The work further advocates labelling of the flagged cryptocurrency wallets and transactions for future transactions. In this way, it will become easier to detect any malicious activity from the flagged accounts in future and by this way one can take timely action and can prevent some catastrophe in future.\\
The other work in the domain of explicitly feature engineering for detection of malicious accounts have been carried out by Farrugia~\textit{et al.} in~\cite{incentive11}. The basic motivation of the work is to figure out the top features which have the largest impact on the outcome of anomaly detection models. Therefore, after thorough examination and evaluation authors identified `Min received value’, `time difference’, and `total balance’ as the three most influential features specifically for Ethereum blockchain. Alongside doing this feature engineering, authors also work over proposing an effective method to detect the malicious accounts, for which authors used XGBoost, which basically is a method of ensemble machine learning via decision tree. 

\subsection{Incentive Attack Classification}
Incentive layer is prone to many attacks ranging from double spending to DDoS attack, etc. \rev{However, the majority of works discussed above focused majorly over either identification of a particular attack,} or identification of some sort of anomalous behaviour in the blockchain model. However, considering the diverse range of attacks, it is equally important for an anomaly detection model to pinpoint the type of attack which is being carried out in the network. \rev{From the perspective of the incentive layer, one such incredible work has been carried out by Sayadi~\textit{et al.} in~\cite{incentive08}. The work focuses on using two separate machine learning} models to first detect the anomalies and then the second to classify it up further. In order to detect outliers in the transactions of blockchain, authors used one class SVM also known as OCSVM, which basically separates novelty outliers on the basis of hyperplane distance among the transactions. Afterwards, the model basically labels them and \final{feeds it} to the next classification model, for which authors used K-means clustering. This K-means clustering is basically a further extension via which authors picked and classified the anomalies into different types of attacks. From the selected anomalies, authors were able to identify the presence of double spending, DDoS, and 51\% vulnerability from the identified labels. \\

\subsection{Market Volatility Prediction}
While talking about incentive or currency layer of blockchain, the aspect of market control, manipulation, and volatility cannot be ignored because it is one of the key aspect over which a huge amount of investment depends upon~\cite{incentivelayer06}. Nevertheless, cryptocurrencies, especially Bitcoin and Ethereum have a large amount of market capitalization and this capitalization is continuously increasing. E.g., the total market capital of Bitcoin in 2017 was \$18 billion, which increased to \$599 billion in 2018~\cite{incentivelayer05}. Considering these aspects, it is important to have answers to know more about these cryptocurrencies, especially about market volatility from an anomalous manipulation viewpoint.\\ 
From a technological point of view, it is important to identify and predict the occurrence of \rev{anomalous factors which can cause huge market volatility. It is also important to predict the occurrence of a major surge in the network because it can also be due to some \final{adversarial attacks} on the network, which can further lead to disastrous outcomes. One such to predict market volatility and return from Bitcoin price movements and transactions movement have been carried out by authors in~\cite{incentive03}. The work highlights that predicting the Bitcoin market is not fairly simple as plenty of complex aspects are associated with this. In order to carry out efficient prediction, authors developed a relationship between the market’s volatility and the complexity measures.} For complexity measures, authors used the concept of transactions connectivity with regard to number of roles, alongside this, authors used the measures from information theoretic perspective as well. Afterwards, these measures were fed into a prediction model, which first characterises the joint behaviour via vector autoregression and afterwards carry out selection and model estimation accordingly.\\
Another interesting work which focuses over identification of market manipulation of Bitcoin due to adversarial and anomalous identities has been carried out by Chen~\textit{et al.} in~\cite{incentive10}. To train and develop the market manipulation model, authors picked a previous database of transaction leakage and organized the transactions into three sub-graphs. Afterwards, authors worked over identification of influence of each account on the fluctuation of the market in order to identify the most influenced accounts, and they carried out this experimentation with the help of singular value decomposition. From SVD, the authors were able to identify certain base accounts and networks, which had a direct relation with the volatility of the network. Similarly, authors were able to identify the types of abnormal transactions which can be carried out between malicious users, e.g., unidirectional, self-loop, bi-direction, polygon, triangle, and star transactions. From the given work, one can efficiently detect the presence of any anomalous factor which can cause market manipulation in the near future. \\
\rev{Market volatility and market manipulation has been in discussion since the advent and popularity of Bitcoin, and a number of graph analysis based works have been carried out so far. However, the integration of other specific anomaly detection models is missing in the literature. Similarly, while discussing market manipulation and volatility in the context of blockchain anomaly detection, it is important to mention that in this section, we only consider works which discuss these aspects from a technological viewpoint.} Contrarily, there are plenty of other works, which purely focus over economical or financial viewpoint, therefore, we did not include these articles in our discussion because they were out of scope of this article. Interested readers can study more about economic growth, volatility, and manipulation of cryptocurrencies in the interesting article written by Bariviera and Sola~\cite{incentivelayer07}.

\subsection{Summary and Insights}
\mubcom{Incentive layer is the major driving force in blockchain technology, which motivates participating nodes to take part in mining and other relevant processes. \rev{Nevertheless, it is a driving force because of the incentivization, but for anomalous peers, it is also one of the most critical layers to target, because they can get direct benefit from this layer in terms of incentives, tokens, etc. Majority of attacks and anomalies over this layer consist of frauds among cryptocurrencies, such as Bitcoin, Ethereum, etc. Apart from cryptocurrency frauds, the second critical anomaly type is \final{financial fluctuations} in the market,} which can cause huge rise and drop among the shares and trading values of assets and currencies over the blockchain network. Another significant direction towards working over anomaly detection from the incentive layer perspective is to identify malicious accounts of the network carrying out such anomalous activities and flag or ban such accounts in order to prevent them for carrying out fraud or market instability. }


\begin{figure*}[]
     \centering

\begin{tikzpicture}


\node [firstblock,  text centered, minimum width = 25em,  text width=25em] (a1) {Anomaly detection in Smart Contract (SC) Layer \\ Sec.~\ref{ContractLayer} };


\node[secondblock, below of=a1, yshift=-3em, xshift=-22em, text width = 4em](b1){Paxos Anomaly on Chain};
\node[secondblock, below of=a1, yshift=-3em, xshift = -16em, text width = 4em ](b2){Faulty Signals \& Tx Detection};
\node[secondblock, below of=a1, yshift=-3em, xshift = -10em, text width = 4em](b3){Anomaly in Contracts Bytecode};
\node[secondblock, below of=a1, yshift=-3em, xshift = -01em, text width = 4em](b4){Detecting Ponzi Contracts};
\node[secondblock, below of=a1, yshift=-3em, xshift = 10em, text width = 4em](b5){Malicious Threats in Contracts};
\node[secondblock, below of=a1, yshift=-3em, xshift = 20em, text width = 4em](b6){Tagging \& Log based Prevention};



\node [fourthblock, below of=b1, xshift = -0em, yshift=-4.5em, minimum width = 2.5mm] (b1c1d1) {\footnotesize{Detecting and Freezing Coins via Novel SC~\cite{contract01}}};

\node [fourthblock, below of=b2, xshift = -0em, yshift=-4.5em, minimum width = 2.5mm] (b2c1d1) {\footnotesize{Application Oriented Anomaly Detection~\cite{contract02,newref09}}};

\node [fourthblock, below of=b3, xshift = -0em, yshift=-4.5em, minimum width = 2.5mm] (b3c1d1) {\footnotesize{Enhancing Bytecode Identification~\cite{contract03, contract07}}};

\node [fourthblock, below of=b4, xshift = -3em, yshift=-4.5em, minimum width = 2.5mm] (b4c1d1) {\footnotesize{Continuous Ponzi Contract Detection\newline~\cite{contract05}}};
\node [fourthblock, below of=b4, xshift = 3em, yshift=-4.5em, minimum width = 2.5mm] (b4c2d1) {\footnotesize{Identified Ongoing Ponzi Schemes on Network~\cite{contract06}}};

\node [fourthblock, below of=b5, xshift = -3em, yshift=-4.5em, minimum width = 2.5mm] (b5c1d1) {\footnotesize{Threats in EVM Contracts~\cite{contract08, contract11, contract12}}};
\node [fourthblock, below of=b5, xshift = 2.5em, yshift=-4.5em, minimum width = 2.5mm] (b5c2d1) {\footnotesize{Efficient Security Threat Identification~\cite{contract09}}};

\node [fourthblock, below of=b6, xshift = -2.5em, yshift=-4.5em, minimum width = 2.5mm] (b6c1d1) {\footnotesize{Updating SC w.r.t Log Pattern~\cite{contract04}}};
\node [fourthblock, below of=b6, xshift = 3em, yshift=-4.5em, minimum width = 2.5mm] (b6c2d1) {\footnotesize{Overhead Reduction in Detection~\cite{contract10}}};


\node[fifthblock, below of=b1c1d1, yshift=-3em, xshift= 0em, text width=4em](z1){Analytical Analysis};
\node[fifthblock, below of=b2c1d1, yshift=-3em, xshift = 0em, text width=4em ](z2){Graph Analysis};
\node[fifthblock, below of=b4c1d1, yshift=-3em, xshift = 0em, text width=4em](z3){Multiple Machine Learning};
\node[fifthblock, below of=b3c1d1, yshift=-3em, xshift = 0em, text width=4em](z4){CNN};
\node[fifthblock, below of=b4c2d1, yshift=-3em, xshift = 0em, text width=4em](z5){SVM};
\node[fifthblock, below of=b5c1d1, yshift=-3em, xshift = 0em, text width=4em](z6){Statistical Analysis};
\node[fifthblock, below of=b5c2d1, yshift=-3em, xshift = 0em, text width=4em](z7){Sequence Learning};
\node[fifthblock, below of=b6c1d1, yshift=-3em, xshift = 0em, text width=4em](z8){Aging Resistant ML};
\node[fifthblock, below of=b6c2d1, yshift=-3em, xshift = 0em, text width=4em](z9){Path Indexing};

\path [line] (a1)-- ($(a1.south)+(0,-0.25)$) -|(b1);
\path [line] (a1)-- ($(a1.south)+(0,-0.25)$) -|(b2);
\path [line] (a1)-- ($(a1.south)+(0,-0.25)$) -|(b3);
\path [line] (a1)-- ($(a1.south)+(0,-0.25)$) -|(b4);
\path [line] (a1)-- ($(a1.south)+(0,-0.25)$) -|(b5);
\path [line] (a1)-- ($(a1.south)+(0,-0.25)$) -|(b6);


\path [line] (b1)-- ($(b1.south)+(0,-0.25)$) -|(b1c1d1);
\path [line] (b2)-- ($(b2.south)+(0,-0.25)$) -|(b2c1d1);
\path [line] (b3)-- ($(b3.south)+(0,-0.25)$) -|(b3c1d1);

\path [line] (b4)-- ($(b4.south)+(0,-0.25)$) -|(b4c1d1);
\path [line] (b4)-- ($(b4.south)+(0,-0.25)$) -|(b4c2d1);

\path [line] (b5)-- ($(b5.south)+(0,-0.25)$) -|(b5c1d1);
\path [line] (b5)-- ($(b5.south)+(0,-0.25)$) -|(b5c2d1);

\path [line] (b6)-- ($(b6.south)+(0,-0.25)$) -|(b6c1d1);
\path [line] (b6)-- ($(b6.south)+(0,-0.25)$) -|(b6c2d1);



\path [line, color=red] (b1c1d1.south)--(z1.north);
\path [line, color=orange] (b2c1d1.south)--(z2.north);

\path [line, color=blue] (b3c1d1.south)--(z4.north);

\path [line, color=brown] (b4c1d1.south)--(z3.north);
\path [line, color=cyan] (b4c2d1.south)--(z5.north);

\path [line, color=violet] (b5c1d1.south)--(z6.north);
\path [line, color=magenta] (b5c2d1.south)--(z7.north);

\path [line, color=olive] (b6c1d1.south)--(z8.north);
\path [line, color=darkgray] (b6c2d1.south)--(z9.north);

\end{tikzpicture}
	\small \caption{Classification of Blockchain Anomalies from Perspective of Contract Layer}
     \label{fig:contractlayer}
\end{figure*}
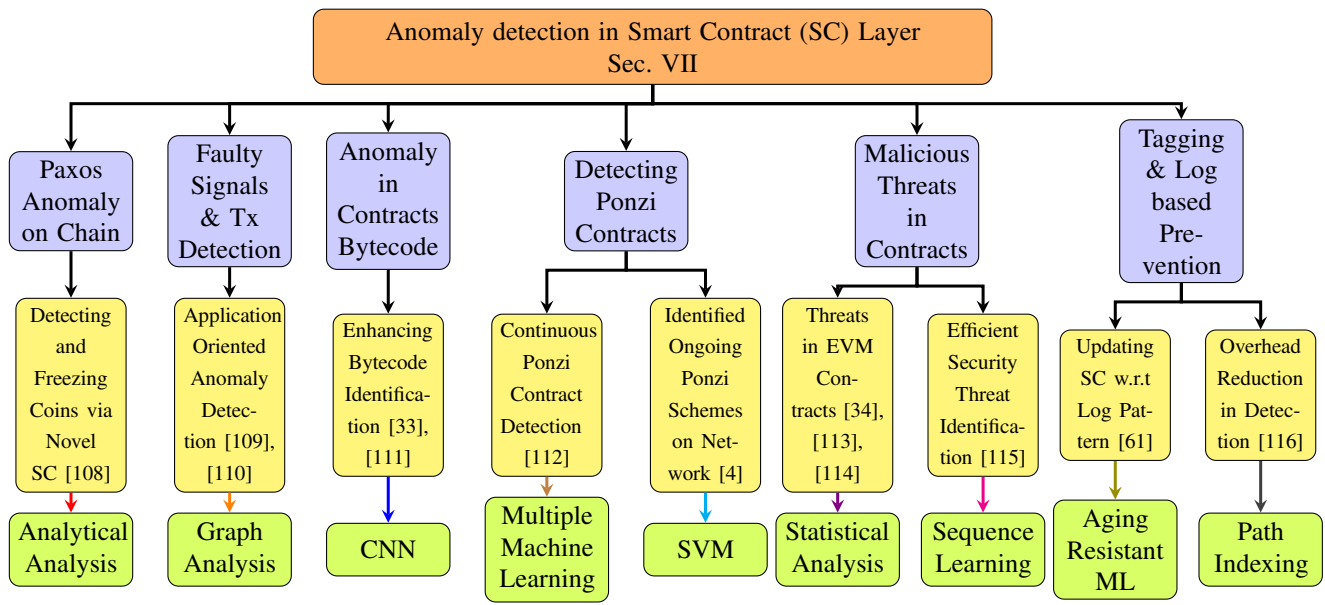



\begin{table*}[ht]
\begin{center}
 \centering
 \scriptsize
  \captionsetup{labelsep=space}
 \captionsetup{justification=centering}
 \caption{\textsc{\\Anomaly Detection in Contract Layer.}}
  \label{tab:contractlayer}
  \begin{tabular}{|P{1.25cm}|P{0.55cm}|P{2.5cm}|P{2.2cm}|P{2.0cm}|P{1cm}|P{1cm}|P{1.8cm}|P{1.1cm}|P{0.6cm}|}
  	\hline
\rule{0pt}{2ex}
\bfseries Domain & \bfseries Ref No. & \bfseries Contribution & \bfseries Detected Anomaly & \bfseries  Anomaly Factors  & \bfseries \centering Blockhain Type &\bfseries Platform \newline Language & \bfseries \comst{Applications} & \bfseries Dataset & \bfseries Compl-\newline exity \\
\hline

\rule{0pt}{2ex}
\centering \textbf{Paxos Anomaly in Blockchain} & \cite{contract01}  & Highlighted a complex immutability related anomaly \& proposed coin freezing. & \tabitem Malicious risky Tx \newline \tabitem Uncommitting Tx & \tabitem Swap frequency \tabitem Tx movement & Public & EVM & \tabitem Digital Assets \newline \tabitem Cryptocurrency & N/S & $-$\\
\cline{2-10}
\hline

\multirow{2}{*}{\parbox{2cm}}
\rule{0pt}{2ex}
\centering \textbf{Faulty Signal and Tx Detection} & \cite{contract02}  & Graphical lasso \& collaborative dictionary learning based anomaly detection in industrial data. & \tabitem Faulty data files & \tabitem Logarithmic loss \newline \tabitem Sample accuracy & Private (Permissioned) & Hyperledger Fabric & \tabitem Digital Assets \newline \tabitem Industrial Data Storage & Real-world Data & $-$\\
\cline{2-10}

\rule{0pt}{2ex}
 & \cite{newref09}  & \mubcom{Graph analysis based anomalous Tx detection from supply chain data.} & \tabitem Faulty unordered transactions & \tabitem Tx life cycle \newline \tabitem Tx order & Private (Permissioned) & Hyperledger Sawtooth & \tabitem Digital Assets \newline \tabitem Supply Chain & Real-world Data & $-$\\
\cline{2-10}
\hline

\rule{0pt}{2ex}
\multirow{2}{*}{\parbox{2cm}{}}

\centering \textbf{Anomaly in Contracts Bytecode} & \cite{contract03}  & Converted bytecode to RGB for efficient anomaly extraction. & \tabitem Compiler bugs in contracts & \tabitem Malicious smart contracts & Public & Solidity & \tabitem Cryptocurrency & Ethereum Contracts & $-$\\
\cline{2-10}

\rule{0pt}{2ex}
& \cite{contract07}  & Identified Malicious Smart Contracts by Assigning Labels on basis of bytecode & \tabitem Anomalous bytecodes & \tabitem Malicious smart contracts & Public & Solidity & \tabitem Cryptocurrency & Ethereum Contracts & $-$\\
\cline{2-10}

\hline

\rule{0pt}{2ex}
\multirow{2}{*}{\parbox{2cm}{}}
\rule{0pt}{2ex}
\centering \textbf{Detecting Ponzi Contracts} & \cite{contract05} & Developed model to predict Ponzi smart contracts from day zero. &  \tabitem Fraudulent smart contract & \tabitem Precision \newline \tabitem Recall \newline \tabitem F-1 Score & Public & EVM & \tabitem Cryptocurrency & Ethereum Data & $-$\\
\cline{2-10}

\rule{0pt}{2ex}
 & \cite{contract06} & Developed model to predict scamming smart contracts in Ethereum. &  \tabitem Malicious Tx \newline \tabitem Malicious SC body & \tabitem Ponzi detection count & Public & EVM & \tabitem Digital Asset \newline \tabitem Cryptocurrency & Real-world Data & $-$\\
\cline{2-10}
\hline

\rule{0pt}{2ex}
\multirow{4}{*}{\parbox{2cm}{}}
\rule{0pt}{2ex}
 & \cite{contract08} & Developed a taxonomy of honeypots of Ethereum smart contracts. &  \tabitem Balance disorder \newline \tabitem Inheritance disorder & \tabitem Hidden traps in SC & Public & Python & \tabitem Cryptocurrency & Ethereum SC data & $-$\\
\cline{2-10}

\rule{0pt}{2ex}
\centering \textbf{Malicious Threats in SC} & \cite{contract11} & Identified 20 Defects in Ethereum smart contracts. &  \tabitem Security,  Availability, Performance, Maintainability, and Re-usability Defects & \tabitem Hidden defects in SC & Public & N/A & \tabitem Cryptocurrency & Ethereum SC data & $-$\\
\cline{2-10}

\rule{0pt}{2ex}
 & \cite{contract12} & Linked Smart Contract Defects to Prospective Unwanted Behaviour. &  \tabitem Contract dependencies  & \tabitem Hidden attacks in SC & Public & Solidity & \tabitem Cryptocurrency & Ethereum SC data & $-$\\
\cline{2-10}

\rule{0pt}{2ex}
& \cite{contract09} & Sequentially learning smart contracts to find weaknesses. &  \tabitem New attack trends & \tabitem Precision \newline \tabitem Recall \newline \tabitem F-1 Score & Public & EVM & \tabitem Cryptocurrency & Ethereum Data & $-$\\
\cline{2-10}
\hline

\rule{0pt}{2ex}
\multirow{2}{*}{\parbox{2cm}{}}
\rule{0pt}{2ex}
\centering \textbf{Tagging \& Log based Anomaly Prevention} & \cite{contract04} & Developed a self-adaptive model to detect log anomaly in smart contracts. &  \tabitem Malicious data storage & \tabitem Time complexity & Public & N/S & \tabitem Cryptocurrency \newline \tabitem Digital assets & Real-world Data & $-$\\
\cline{2-10}

\rule{0pt}{2ex}
& \cite{contract10} & Detected \& prevented abnormal control flow in Ethereum smart contracts. &  \tabitem Control paths & \tabitem Gas consumption \newline \tabitem Overhead & Public & EVM & \tabitem Digital assets \newline \tabitem Cryptocurrency & Multiple datasets & $-$\\
\cline{2-10}
\hline

 \end{tabular}
  \end{center}
\end{table*}


\section{Anomaly Detection in Contract Layer of Blockchain} \label{ContractLayer}

In this section, we provide a detailed review of existing works from the perspective of anomaly detection in contract layer of blockchain (cf. Section~\ref{BlockLayers} for details)~(cf. Fig.~\ref{fig:contractlayer} and Table.~\ref{tab:contractlayer}). 

\subsection{Paxos Anomaly in Blockchain}
As the name suggests, in this section we will be discussing an anomaly related with dependent transfers, which is categorized under the name of the famous consensus protocol `Paxos’. In order to understand this anomaly a bit further, it is important to understand two major concepts, one is the concept of Paxos consensus and the other is termination of a consensus model in a decentralized blockchain environment. \rev{From the perspective of Paxos, it can formally be defined as a family of selected protocols which can be used to reach a consensus in an unreliable processor network~\cite{contractlayer08}.}\\ 
Moving towards discussion of similarities between Paxos anomaly and blockchain anomaly, it is important to highlight a critical and thorough work carried out by Natoli and Gramoli in~\cite{contract01}. \rev{Authors provided a thorough analysis of how the asynchronous nature of blockchain and message delays can cause a major issue in termination of consensus, which can then lead to the start of two simultaneous chains} where both the miners agrees simultaneously over their own `$k$’ set of blocks. Authors further evaluated that this problem accelerates and becomes more catastrophic in case of dependent transactions especially in case of a private blockchain, which can further lead to double spending attack and uncommitting of transactions. Authors further developed a complete model to study this effect in which they evaluated by automating the anomaly reproduction in the decentralized network at different difficulties involved in mining the block. Through the evaluation, authors show the hazardous effects which can be caused if proper actions are not taken. Finally, authors discussed a prospective smart contract based solution in which authors provided certain examples showing the specific conditions and statements which can be added in a smart contract to overcome the occurrence of this anomaly, e.g., detecting and freezing of coins at the time of need. Authors further emphasized that modern blockchain networks, such as Ethereum, etc should develop more secure smart contracts so that the possibility of such anomalies can be eradicated completely. \rev{It will not be wrong to say that Paxos is one specific type of anomaly that freezes the transactions on the basis of dependency, however, with more in-depth analysis, one can work over identification of other similar anomalies in the blockchain network. }

\subsection{Faulty Signals \& Tx Detection}
Researchers are actively working in development of modern blockchain systems, which are paving paths towards development of `Blockchain 3.0’~\cite{contractlayer09}. However, this also comes up with affiliated anomalous concerns, such as malicious attacks and faulty signals and transactions. In order to overcome this anomaly detection issue, authors in~\cite{contract02} worked over proposing a deterministic smart contract based anomaly detection in blockchain based IoT networks. Authors work \final{over the} development of a collaborative learning approach in which authors used the functionalities of \rev{probabilistic dictionary learning to figure out the existence of a particular faulty signals anomaly in the blockchain based IoT network. To carry out the work, the authors first formulated the problem by developing an anomaly score on the basis of clients, network participants, and used dataset.} Afterwards, authors developed the algorithm for density estimation in the blockchain network in a collaborative manner, which further led to the development of protocol which is used to update the parameters data. From the evaluated models, authors claimed that the proposed model \rev{identified anomalies in a more accurate manner as compared to the \final{previous works.}\\
\mubcom{Another similar work from the perspective of anomaly detection over supply chain network operating over Hyperledger Sawtooth has been carried out by Oh~\textit{et al.} in~\cite{newref09}. The work focuses on identification of anomalies via graph analysis, which is done with the help of a smart contract. The smart contract is designed in such a manner that it identifies whether the order of a transaction is correct, and it has all the necessary linkage with its predecessors. If the transaction obeys all the conditions, then it is recorded over the ledger,} elsewise, it is flagged as an anomalous transaction. Authors successfully implemented this anomaly detection notion over Hyperledger Sawtooth in the form of an additional layer, which can be integrated with blockchain platforms to ensure the capability of anomaly detection. From the presented work, it can be seen that detecting faulty signals is important and it can also be seen in future applications of blockchain. However, the detection phenomena are not much advanced, therefore, it is equally important to detect the occurrence of these faulty signals in order to take appropriate actions.}

\subsection{Anomaly in Contracts Bytecode}
With the development of Ethereum smart  contracts, a plethora of \rev{opportunities and future directions which require the integration of decentralized services for their efficient functionalities can be visualized.} Since, smart contracts is a deterministic piece of code which cannot be stopped once it starts execution on the blockchain network, therefore, it is equally important to ensure that the outcome of a particular smart contract is in the favour of blockchain network, and it will not cause any catastrophe, which is done with the help of anomaly detection models. 
One such work towards identification of anomalies in bytecode of Ethereum smart contracts has been carried out by Huang in~\cite{contract03}. Unlike other similar works, Huang did not focus primarily on extraction of novel features for efficient identification, instead, the major focus of the article is to reduce the overall labour cost associated with identification of anomalies in the Ethereum bytecode. In order to do so, authors first work over translation solidity bytecode into an RGB code, which is further used to develop an encoded image of the fixed-size. The RGB image is then fed to CNN for training, which automatically extract features, carry out learning, and then carry out detection of \rev{bugs of the compiler at the time of execution of smart contract.} \final{In this way,} the proposed work is able to identify bugs in a more cost effective and efficient manner as compared to previous works. Another critical work to enhance bug prediction accuracy for smart contracts has been carried out by Kim~\textit{et al.} in~\cite{contract07}. Authors worked over analysing smart contract bytecode in order to categorize and attribute them in the form of tags for swift identification. In order to do so, \rev{the authors used a learning model comprising five different stages ranging from pre-training stage to inference stage.} In order to evaluate the proposed methodology, authors used code examples EtherScan and Google BigQuery datasets. From the outcome results, it can be seen that the authors were able to successfully classify smart contracts \rev{bytecode on the basis of attributes present in them. \rev{Since bytecode is being used in a large number of blockchain simulators, therefore, more research work needs to be carried out over detection of anomalies in bytecode for smooth running of the network.}}

\subsection{Detecting Ponzi Contracts}
Since the advent of blockchain and cryptocurrencies, adversaries are continuously trying to take unusual and illegal advantages of \rev{certain hidden functionality of it which common people are not aware of. One such type of fraudulent model is the development of the Ponzi scheme on the decentralized blockchain network, which not only affects an individual, but it also affects the economy on a deeper level~\cite{contractlayer10}.} 
An interesting work towards development of data mining models to develop Ponzi smart contracts on Ethereum blockchain has been carried out by Jung~\textit{et al.} in~\cite{contract05}. Authors first highlighted the functioning and basic methods being used in Ponzi smart contracts by specifically focusing over Ethereum, and afterwards authors worked over building a dataset of Ponzi contracts on the network. Then the authors used these malicious Ponzi contracts to pick out specific features, for which authors used the transactions and the compiled code on the network for these malicious contracts. \rev{Then authors work over \final{the development} of a classification model, which efficiently predicts the presence or absence of malicious and Ponzi factors in a smart contract.} Authors carried out evaluation of their proposed model for 250 days and the outcome results identified that it predicts malicious contracts in an efficient manner. Another similar work focusing over, exploitation of Ethereum blockchain to identify Ponzi contracts has been carried out by Chen~\textit{et al.} in~\cite{contract06}. In order to diversify their search, authors first manually picked 200 Ponzi smart contracts by analysing around 3,000 available Ethereum contracts. \rev{After that, authors extracted two malicious features on the basis of operation codes and history of transactions. Afterwards, authors used data mining tools to develop the complete model which classified each new smart contract as Ponzi or safe.} From the analysis, authors highlighted that more than 500 Ponzi schemes are currently being operated on the blockchain network.\\
\rev{It is important to mention that the majority of the work towards development of Ponzi smart contracts has been carried out from Ethereum perspective because they are the first ones to introduce the feature of smart contracts in blockchain, therefore, they are the most vulnerable one. However, these Ponzi schemes are not just limited to Ethereum, as they are spreading to other cryptocurrencies and blockchains as well. Therefore, there is a dire need to develop such advanced detection \final{models which} accurately analyse, detect, and eradicate such Ponzi smart contracts before occurrence of any catastrophe. }

\subsection{Malicious Threats in Smart Contracts}
\rev{Every smart contract being executed in a blockchain network has its own dependencies and can affect the blockchain network in its own way. Similarly, once executed, it is impossible to stop the functioning of smart contracts.} Therefore, certain time adversaries take unfair advantage of this feature and try to add certain malicious threats and honeypots in smart contracts which can cause a serious harm to the network or individual. Therefore, destruction and timely identification of such smart contracts is mandatory to keep the network safe from adversaries. One such work towards identification of malicious honeypots on Ethereum smart contracts has been carried out by authors in~\cite{contract08}. The authors developed a tool for honeypot identification and named the tool as ‘HoneyBadger’. \rev{To elaborate their concept a bit further, authors proposed a formal definition of \final{honeypots in} which they described a honeypot in a specific type of smart contract which tricks users to give their funds to attackers in the exchange of some leaked arbitrary funds. In order to attract an audience, the attacker first deploys a smart contract which seems to be giving funds to the executor. Then, the victim falls prey to the greed of getting more funds, and thus he/she transfers the required sum to the attacker. Finally, the attacker withdraws both the funds and the original funds, and the victim is left with nothing in hand.} Through visual examples, authors explained the severity of the situation and thus to overcome this, authors developed a complete taxonomy of such honeypots which are currently running over the Ethereum network. The authors also carried out an extensive analysis of such honeypots in contacts on the basis of their sub-components, such as various disorders and overflows. \\
\rev{Another critical work working over identification of critical defects in the smart contracts of Ethereum has been carried out by Chen~\textit{et al.} in~\cite{contract11}.} Authors developed the motivation of their work by discussing that certain smart contracts can have defects, and some severe defects can deeply affect the functioning of the whole network and can impact the whole chain. Afterwards, authors identified defects in contracts by analysing gas consumption, keywords filtering, open card sorting, and similar other features. In this way, authors were able to successfully identify 20 critical defects which can cause severe issues in the network. The authors further classified these contracts to five subtypes named as security defects, availability defects, performance defects, maintainability defects, reusability defects. As an extension of this work, the authors proposed a complete tool and named it as DefectChecker~\cite{contract12}. \rev{The proposed tool can detect 08 critical defects in the malicious contracts which could have caused abnormal and unwanted behaviour. Afterwards, they worked over using the tool} for the identification of level of impact. From the outcomes and experimental results, it can be seen that the proposed model can predict the given contracts with 88.8\% F-Score. In this way, the authors were able to conclude that out of 1,65,621 analysed smart contracts, 25,815 had at least one identified as defective.\\
Till now, the works used analytical and statistical modelling and analysis to identify prospective threats in the smart contracts. However, a detailed model using sequence learning to carry out similar work has been presented by Tann~\textit{et al.} in~\cite{contract09}. In order to make the identification effective, authors worked over using long short term memory (LSTM machine learning model. For which, authors first classified the threats of smart contracts and then sequentially modelled them on the basis of opcode sequence. Then after labelling the data through \textit{Maian}, authors used supervised learning to predict the smart contracts having critical threats. In this way, authors were able to identify threats with 99.57\% test accuracy. 

\subsection{Tagging \& Log based Anomaly Prevention}
The integration of smart contract technology with decentralized blockchain network has initiated a new era of decentralized on-chain agreement. \rev{Due to this initiation a large number of applications are now being developed which utilize the tremendous advantages of smart contracts. But, this also comes up with the associated risks and concerns which needs to be eradicated in a timely manner.} One such work using log systems for smart contracts to identify prospective anomalies has been carried out by Shao~\textit{et al.} in~\cite{contract04}. \final{Authors proposed an LSC architecture, via which authors formulated a} complete framework which can detect anomalies by users with the help of efficient smart contracts. The protocol works over learning and analysing logs on the basis of aging-resistant machine learning models. Afterwards, the learnt output results which can also be used as models of anomaly detection are forwarded to executable smart contracts in order to identify presence of anomalies in the network. The developed smart contract also keeps on updating on the basis of new available information in order to ensure the novelty and security of smart contract against vulnerabilities. \rev{Another interesting work towards usage of tagging systems to defend smart contracts of Ethereum have been carried out by Wang~\textit{et al.} in~\cite{contract10}. The major goal of the work is to prevent execution of malicious smart contracts alongside enhancing the overhead of detection. In this way, authors can ensure that all nodes, even with a small execution power will be able to run the contract} without worrying about the overhead. From the empirical analysis, authors showed that their proposed model can effectively safeguard against 11 specific errors and attacks such as logic error, superficial randomness, abnormal control flow, etc. Authors further evaluated their proposed model to identify whether the proposed model is practical or not, and from the experimental results, it can be concluded that the proposed ContractGuard model only causes an additional 28.27\% runtime overhead and 36.14\% deployment overhead. \rev{After analysing the works, it can be mentioned that the applications utilizing tagging \& log based design are increasing, similarly, the anomalies over these applications are also increasing. Therefore, there is a dire need to develop machine/deep learning based models which can detect the occurrence of these anomalies within an actionable time frame.}

\subsection{Summary and Insights}
\mubcom{Contract layer is relatively a more technical and considerably a new layer in blockchain which got famous in the second era of \final{blockchain,} named as blockchain 2.0 when Ethereum platform provided its users the functionality of developing DApps. A large number of contracts in a well-established blockchain network are pre-developed and do not contain any bugs, however, the malicious participants in the network always try to find out loopholes by any means and smart contracts are their recent targets because it is hard for a non-technical person to identify bugs and honeypots in the smart contracts. The work from the perspective of detection of anomalies in this specific layer is divided into multiple types ranging from identification of contracts restricting dependent transactions to highlighting faulty signals being transmitted via deployment of a smart contract. However, the most prominent works in the anomaly detection over this layer have been carried out from perspective of detection of Ponzi schemes and detection of critical security threats in the contracts, such as hacking, etc.}

\section{Challenges \& Future Research Directions}

\subsection{\rev{Data Layer}}

\subsubsection{Privacy Preserving Anomaly Detection in Blockchain}
\paragraph{Key Challenge} From our analysis of blockchain based anomaly detection models, we observed that none of the work discussed integration of privacy preservation in their works. Nevertheless, blockchain works over the phenomenon of a decentralized ledger and every node has a copy, therefore, it has got a lot of privacy issues that researchers are tackling~\cite{survey12}. Similarly, from the perspective of anomaly detection in blockchain, this issue doubles because one needs to analyse even the \rev{deep details of each transaction on the data ledger of blockchain in order to identify any anomalous behaviour.} \rev{However, it is highly unwilling that blockchain participants share their complete data unless they have been provided with a complete privacy guarantee.}
\paragraph{Future Directions} Considering the nature of privacy requirement in blockchain based anomaly detection, it will not be wrong to say that there is a dire need to work over this issue. \rev{In order to overcome such issues, researchers can work over integration of modern privacy preservation strategies, such as differential privacy~\cite{futureref01}, zero knowledge proofs~\cite{futureref02}, etc. with anomaly detection models of blockchain.} In this way, researchers will be able to provide blockchain users with a safe and secure platform via which they will be able to prevent prospective anomalies without the risk of losing their private data. It is important to highlight that each of the privacy preservation models comes with a take away, e.g., while employing differential privacy, one has to deal with a trade-off between utility and privacy, same goes with other privacy preserving models. \rev{Therefore, the prospective models which show minimum effect over the utility and \final{privacy leakage} of the system will be a key contribution in this domain of privacy preserving anomaly detection in blockchain technology.}

\subsubsection{Integrating Federated Learning with Blockchain Anomaly Detection}

\paragraph{Key Challenge} \rev{In order to make efficient anomaly detection decisions, one has to train their machine learning model in a centralized manner. Similarly, in the recent years,} researchers worked over integration of various machine learning based anomaly detection for blockchain technology including CNN, SVM, etc. However, as per our observation, none of the work has integrated federated learning for anomaly \final{detection in blockchain.} Nevertheless, blockchain is a decentralized model and the basic phenomenon of federated learning is also leading in a decentralized manner instead of a centralized server. Therefore, these two technologies perfectly fit with each other from the perspective of framework. \rev{Similarly, certain works have also identified the effectiveness of federated learning in anomaly detection of IoT and similar technologies~\cite{futureref03, futureref04}.} Now, the need is to develop such federated learning based anomaly detection models which comply with the nature of blockchain technology.

\paragraph{Future Directions} Integrating federated learning with blockchain anomaly detection has two fold advantages. One from the perspective of security and trust in the network, and the other \rev{from the perspective of reduction of computational overhead and data storage.} Federated learning already has decentralized nature, therefore, anomaly detection models do not have to collect huge amount of data in centralized servers, which will enhance and prevail a sense of trust in the network and blockchain. Similarly, from the second viewpoint, it is important to mention that detection overhead and data storing in a \rev{centralized database are the two critical issues \rev{which the majority of  anomaly detection models are facing.} However, if an anomaly detection model starts working in a decentralized manner,} then these major issues can be reduced to a negligible level, which can be done with the help of federated learning. 
\subsection{\rev{Network Layer}}

\subsubsection{Development of Efficient Consensus Models for Anomaly Detection}
\paragraph{Key Challenge} Since the advent of blockchain, a vast number of consensus mechanisms are being developed by researchers and experts to enhance the aspect of trust among peers and to overcome any prospective vulnerability in the network~\cite{futureref08}. As in the consensus model, all nodes reach consensus over a unified transaction, similar to this, in case of a detected \final{anomaly, all} nodes have to reach consensus that the nominated vulnerability is an anomaly. This becomes even more difficult when a vulnerability is not universally recognized or identified as an anomaly. E.g., for some nodes, an anomaly can just be a random behaviour but for other nodes, it could be a point of deep concern. Therefore, in such cases, reaching a consensus on a unified opinion becomes even more difficult. \rev{This can be done by developing efficient blockchain oriented network communication models, which will facilitate blockchain nodes to reach a unified consensus over anomalies as well.}

\paragraph{Future Directions} \rev{It has been proven that the consensus carries an importance of backbone in blockchain technology, because blockchain nodes can reach and agree upon a unified claim due to this feature.} Similarly, in case of anomaly detection, this consensus needs to be finalized in a deterministic way so that none of the adversaries take unusual advantage of it. However, till now, as per our knowledge, there is no specific communication and consensus model which facilitates the early finality of consensus in case of an anomaly. Therefore, there is a strong need to develop such models, which have specific features regarding detection of an anomaly. \rev{This can also be done by integrating some specific functionalities of anomaly detection in current running consensus models. E.g., a specific feature can be enabled if an anomaly is reported by a trustworthy mining node, then a specific network \final{protocol can be used to} disseminate quickly, or in case of an anomaly detection via some detection model, dissemination should be done via a specific protocol, etc.} Similarly, certain penalty functions can be formulated and can be added in the existing consensus models, which reward or penalize the reporting of true or false anomalies. Nevertheless, this field of consensus modification in blockchain technology is pretty huge and it has a large gap especially for anomaly detection oriented consensus models which can be explored by researchers. 

\subsection{\rev{Incentive Layer}}
\subsubsection{\rev{Market Manipulation Detection}}
\paragraph{\rev{Key Challenge}} \rev{Market manipulation in blockchain based applications can be defined as a targeted attempt to influence the market price of an asset/cryptocurrency in an artificial manner~\cite{comstref13}. This is usually carried out by a single participant or a group of participants who aim to create an artificial illusion about an asset or cryptocurrency in the corresponding market in order to fluctuate the price. The actual aim of these participants is to get the profit from the end result of these fluctuations. One of the most famous market manipulation strategy is pump and dump, in which an individual or a number of people from a specific class/group try to pump the value of a penny stock (a stock whom value is less than one dollar) by spreading a plethora of fake news about it with an aim to get the profit once it reaches its peak value~\cite{comstref15}.}

\paragraph{\rev{Future Direction}}
\rev{The concept of market manipulation is not new, and it has been there since the start of financial markets. However, with the advent of decentralized digital markets for assets and cryptocurrencies, this aspect has taken a different turn and malicious bodies continuously try to manipulate the market to get unfair advantage. Therefore, time detection and notification about any prospective or present market manipulation is extremely important in order to save the community from losing their valuable assets. Nowadays, the modern machine learning based anomaly detection models are being used to detect and predict the fluctuations and manipulations in the market. One such example is the use of classification based supervised machine learning models in order to learn \final{from the data} and predict the occurrence of any prospective manipulation~\cite{comstref14}. However, it will not be wrong to say that this field still requires considerable attention from research, academia, and industry, in order to develop more advanced detection models. For example, unsupervised machine learning models can be developed for such cryptocurrencies and assets, who do not have a substantial amount of data to train models. Similarly, models need to be developed in accordance with modern market manipulation strategies, such as wash trading, whale wall spoofing, etc.}

\subsection{\rev{Contract Layer}}

\subsubsection{Malicious Threats Identification in Modern Smart Contract Platforms}

\paragraph{Key Challenge} \rev{Ethereum introduced the usage and functionality of smart contracts in blockchain technology.} However, now almost every new blockchain model has its own smart contracts for its functioning. One of the largest examples after Ethereum is Hyperledger platform, which has its own diverse range of smart contracts especially focusing on enterprise functioning~\cite{futureref05}. Similarly, all other blockchain platforms have their own personalized smart contract which facilitates their functions. However, if we analyse the integration of anomaly detection in blockchain smart contracts, \rev{the majority of the work just focuses on anomaly detection in Ethereum smart contracts. No doubt, Ethereum was the first one to introduce smart contracts, therefore, there is a huge amount of literature on it. But now there is a need to work over identification of anomalies in smart contracts from other technologies as well.}
\paragraph{Future Directions} From our analysis regarding integration of anomaly detection blockchain for malicious smart contract identification, the majority of work we found only targets Ethereum based smart contracts. Considering the recent development of blockchain based technologies, we believe integration of anomaly detection with smart contracts of other technologies can be a key towards development of secure blockchains. For example, Hyperledger Fabric is one of the most viable alternatives to Ethereum, however, a very minimal literature highlighting anomalous effects in Hyperledger Fabric can be found. Similarly, certain other blockchain platforms, such as Stellar, Waves, Nem, etc. have tremendous smart contract features, however, no work is available over identification of anomalous users and contracts for these technologies. It is important to highlight that while developing anomaly detection models for new blockchain \final{models, one} key thing that needs to be kept in mind is their community standard. E.g., some communities might be willing to share significant amounts of information towards the cause of anomaly detection. Contrarily, some communities might be stricter in sharing the data for development and execution of anomaly detection models. Therefore, while developing such models, the aspect of community requirements and standards needs to be taken into consideration. 

\subsection{\rev{Cross-Layers}}

\subsubsection{\rev{Rapidly Evolving Attacks over Blockchain Network}}
\paragraph{\rev{Key Challenge}} \rev{Blockchain attackers are also getting strong and are continuously trying to figure out novel attack surfaces for the blockchain network. One of the most prominent types of attack among these rapidly evolving attacks is cross-layer attacks. As the name suggests, the aim of these types of attacks is to target multiple layers or to utilize the functionality of one layer to target the operations at another layer. For example, in case of DNS resolution attack, an attacker trying to poison the cache of domain name server over the network layer, which results in malicious query results for IP related queries~\cite{comstref09}. In this way, the malicious node tries to trick the network into recording fake transactions or blocks over the data layer of the blockchain. Similarly, in case of BGP attack on blockchain network, the server works over diverting and re-routing the blockchain traffic by publishing fake announcements in the network. In modern blockchain networks, this can be done by initiating a malicious smart contract over the network, which contains such clauses. Similarly, in case of smart contract denial of service (DoS) attack, the network layer is targeted with the help of malicious smart contracts, which results in slowing down the processing time of transactions and blockchain in order to reduce the transaction rate~\cite{comstref10}.}
\paragraph{\rev{Future Directions}}
\rev{The recent developments in blockchain attacks, especially the cross-layer attacks opened a wide-attackable surface for malicious users, where they can take unfair advantage of utilization of one layer for attacking another layer.} \revtwo{Therefore, it is equally important to develop advanced mechanisms in order to detect and tackle them before harming the network.} For example, one of the solutions of DNS resolution attack is to carry out routing-awareness~\cite{comstref11, survey09}. But in order to tackle it, it is equally important to detect the occurrence of attack within actionable time, which can be done by developing advanced anomaly detection models. Similar is the case with other rapidly evolving attacks, where timely detection and action is required. \rev{In case of cross-layer attacks, unsupervised learning based anomaly detection approaches can play a vital role, because these models are capable of detecting anomalies without available training data. Thus, research works can be carried out in advancement and development of such anomaly detection models in order to overcome the risk of rapidly evolving attacks over the network.}

\subsubsection{\rev{Anomalous Attacks During Interoperable and Distinct Blockchain Networks}}
\paragraph{\rev{Key Challenge}} \rev{Each blockchain network has its own unique features and functionalities, which do not overlap with others. For example, one blockchain might only require the selected validators to validate the block, contrarily, the other would require strong PoW. Similarly, an important issue in this context that these platforms face is their interoperability, which basically refers to the phenomenon that allows two independent blockchain networks to carry out exchange of data and information without each other. Nevertheless, it is a progressive direction towards development of modern blockchain infrastructure where different blockchains will be able to exchange information with each other without any dependency, but this comes with its own unique set of challenges. For example, a specific blockchain network working over PoW consensus model (which has its own associated anomalies) is going to interact with another blockchain network working over PoS consensus model (having its own attack surface for anomalies). \rev{This interaction on one hand facilitates interoperability, but on the other hand it also increases the attack surface for malicious attackers.} Therefore, there is a need to develop such anomaly detection models, which are suitable for multiple blockchains alongside their interoperable domain.}
\paragraph{\rev{Future Direction}}
\rev{As we improve the types of architectures and their interoperability, the vulnerable attack surface also increases. Therefore, it is equally important to develop adequate defence mechanisms to overcome these attacks. This process of overcoming attacks basically starts with the development of efficient attack/anomaly detection models. \rev{Each blockchain has its own operations and may have a well-suited anomaly detection model for itself, but there is a strong possibility that one specific anomaly detection model (which is suitable for that specific blockchain) might not \final{work with similar accuracy over} other blockchain networks. Therefore, in order to enhance the performance interoperability among multiple blockchain networks, it is important to design such flexible and dynamic anomaly detection models, which also work with dynamic blockchain environments.}}

\section{Conclusion}
Since the beginning of blockchain technology, it has attracted attention from both academia and industry. One of the prime reasons behind this attention is the P2P architecture of \final{blockchain which} makes it a secure, trustworthy, and truthful platform which is immutable and can be verified at the time of need. \rev{Even though blockchain has benefits, it is also vulnerable to a number of attacks by adversaries, such as security, privacy, reliability, and performance attack, etc. Therefore, in order to keep these functionalities in full running condition,} it is important to identify any anomalous behaviour in the network within a limited time. In order to do so, anomaly detection techniques come into effect, which identify any anomalous behaviour in the network and report it for timely action. In this article, we work over providing a thorough survey of these anomaly detection models. Firstly, we provide a \rev{thorough discussion of how anomaly  detection can help in enhancing the trust and security of the blockchain network and its ongoing applications.} Afterwards, we provide a detailed classification of blockchain anomalies alongside discussing evaluation \final{metrics and} key requirements for the development of anomaly detection models in the network. Afterwards, we provide a detailed in-depth analysis of existing anomaly detection works from the perspective of four most prominent layers of blockchain technology. Finally, we provide a comprehensive discussion about certain challenges and future research directions which needs attention from the researchers working in the field of anomaly detection in blockchain technology.

\bibliographystyle{IEEEtran}

\begin{IEEEbiography}[{\includegraphics[width=1in,height=1.25in,clip,keepaspectratio]{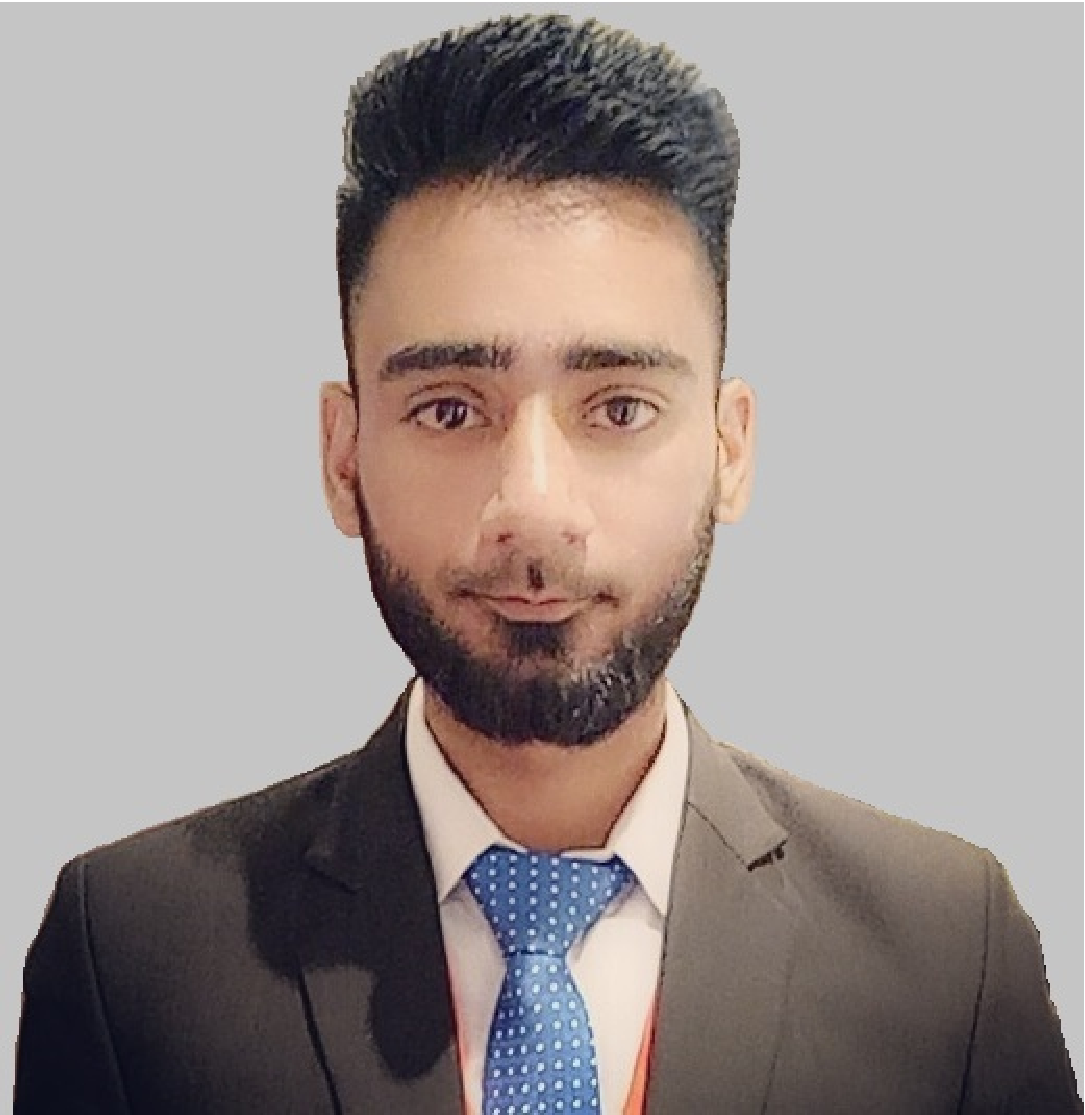}}]{Muneeb Ul Hassan} received his Ph.D. degree from Swinburne University of Technology, Hawthorn VIC 3122, Australia. He received his Bachelor degree in Electrical Engineering from COMSATS Institute of Information Technology, Wah Cantt, Pakistan. He received Gold Medal in Bachelor degree for being topper of Electrical Engineering Department. His research interests include privacy preservation, blockchain, smart grid, and anomaly detection.

\end{IEEEbiography}

\begin{IEEEbiography}[{\includegraphics[width=1in,height=1.25in,clip,keepaspectratio]{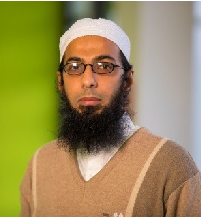}}]{Mubashir Husain Rehmani (M’14-SM’15)} received the M.S. degree from the University of Paris XI, Paris, France, in 2008, and the Ph.D. degree from the University Pierre and Marie Curie, Paris, in 2011. He is currently working as Lecturer at Munster Technological University (MTU), Ireland. He has been selected for inclusion on the annual Highly Cited Researchers™ 2020 and 2021 list from Clarivate. 

\end{IEEEbiography}

\begin{IEEEbiography}[{\includegraphics[width=1in,height=1.25in,clip,keepaspectratio]{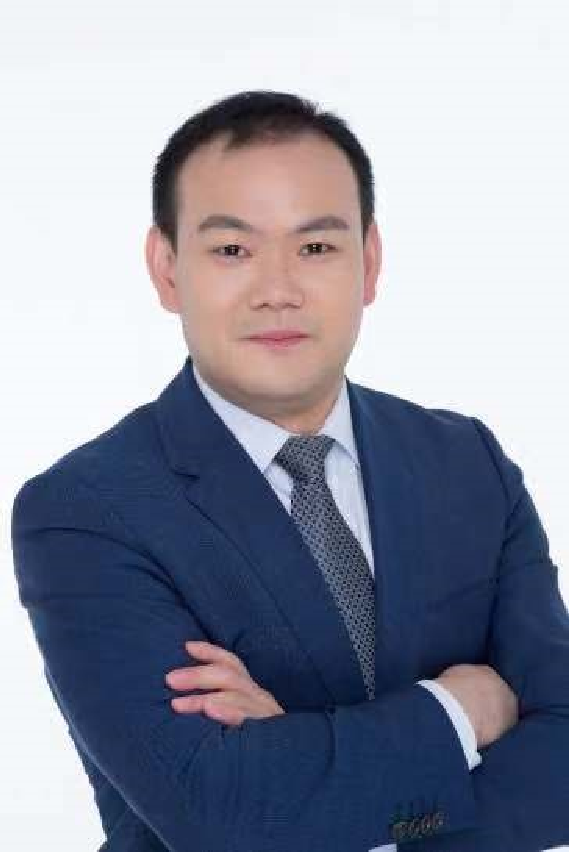}}]{Dr. Jinjun Chen} is a Professor from Swinburne University of Technology, Australia. He is Deputy Director of Swinburne Data Science Research Institute. He holds a PhD in Information Technology from Swinburne University of Technology, Australia. His research results have been published in more than 160 papers in international journals and conferences, including various IEEE/ACM Transactions. 
\end{IEEEbiography}

\end{document}